\documentclass{aa}

\usepackage{hyperref}
\usepackage{txfonts}

\usepackage{wasysym}

\newcommand{\mum} {$\muup$m}
\newcommand{\MJup}{$M_\mathrm{Jup}$}
\newcommand{\RJup}{$R_\mathrm{Jup}$}
\newcommand{\teff}{$T\!_\mathrm{eff}$}
\newcommand{\logg}{$\log g$}
\newcommand{\Ms}{$M_\mathrm{s}$}
\newcommand{\Ks}{$K_\mathrm{s}$}
\newcommand{\Lp}{$L^\prime$}
\newcommand{\Bralpha}{Br$\alphaup$}
\newcommand{\roxsB}{ROXS\,42B}
\newcommand{\roxsBb}{ROXS\,42B\,b}

\begin{document}

\title{Mid-infrared characterization of the planetary-mass companion ROXs~42B~\MakeLowercase{b}}

\titlerunning{\roxsBb}
\authorrunning{Daemgen et al.}

\author{Sebastian Daemgen\inst{\ref{inst1}} 
\and Kamen Todorov\inst{\ref{inst1},\ref{inst2}} 
\and Jasmin Silva\inst{\ref{inst3}} 
\and Derek Hand\inst{\ref{inst4}} 
\and Eugenio V. Garcia\inst{\ref{inst4},\ref{inst5}} 
\and Thayne Currie\inst{\ref{inst6}} 
\and Adam Burrows\inst{\ref{inst7}} 
\and Keivan G. Stassun\inst{\ref{inst8}} 
\and Thorsten Ratzka\inst{\ref{inst9}} 
\and John H. Debes\inst{\ref{inst10}} 
\and David Lafreniere\inst{\ref{inst11}} 
\and Ray Jayawardhana\inst{\ref{inst12}} 
\and Serge Correia\inst{\ref{inst13}}
  }

\institute{
ETH Z\"urich, Institut f\"ur Astronomie, Wolfgang-Pauli-Strasse 27, 8093 Z\"urich, Switzerland, \email{daemgens@phys.ethz.ch}\label{inst1}
\and Anton Pannekoek Institute for Astronomy, University of Amsterdam, Science Park 904, 1098 XH Amsterdam, Netherlands\label{inst2}
\and Department of Physics and Astronomy, University of Hawaii-Hilo, 200 W Kawili St, Hilo, HI 96720, USA\label{inst3}
\and Lawrence Livermore National Laboratory, 7000 East Ave, Livermore, CA 94550, USA\label{inst4}
\and Lowell Observatory, 1400 West Mars Hill Road, Flagstaff, USA\label{inst5}
\and National Astronomical Observatory of Japan, Subaru Telescope, Hilo, HI, USA\label{inst6}
\and Astrophysical Sciences, Princeton University, 4 Ivy Lane, Princeton, New Jersey 08544, USA\label{inst7}
\and Department of Physics and Astronomy, Vanderbilt University, Nashville, TN 37235, USA\label{inst8}
\and Institute for Physics/IGAM, NAWI Graz, University of Graz, Universit\"atsplatz 5/II, 8010 Graz, Austria\label{inst9}
\and Space Telescope Science institute, Baltimore, MD 21218, USA\label{inst10}
\and Institut de Recherche sur les Exoplan\`etes et Universit\'e de Montr\'eal, Montr\'eal, QC, Canada\label{inst11}
\and Faculty of Science, York University, 4700 Keele Street, Toronto, ON M3J 1P3, Canada\label{inst12}
\and Institute for Astronomy, 34 'Ohi'a Ku St. Pukalani, HI\label{inst13}
}

\abstract{
  We present new Keck/NIRC2 3--5\,\mum\ infrared photometry of the planetary-mass companion to \roxsB\ in \Lp, and for the first time in Brackett-$\alpha$ (\Bralpha) and in \Ms-band. We combine our data with existing near-infrared photometry {and $K$-band (2--2.4\,\mum) spectroscopy} and compare these data with models and other directly imaged planetary-mass objects using forward modeling and retrieval methods in order to characterize the atmosphere of \roxsBb. The  \roxsBb\ 1.25--5\,\mum\ spectral energy distribution most closely resembles that of GSC\,06214\,B and $\kappa$\,And\,b, although it has a slightly bluer \Ks$-$\Ms\ color than GSC\,06214\,B and thus currently lacks evidence of a circumplanetary disk. We cannot formally exclude the possibility that any of the tested dust-free/dusty/cloudy forward models describe the atmosphere of \roxsBb\ well. However, models with substantial atmospheric dust/clouds yield temperatures and gravities that are consistent when fit to photometry and spectra separately, whereas dust-free model fits to photometry predict temperatures/gravities inconsistent with the \roxsBb\ $K$-band spectrum and vice-versa. Atmospheric retrieval on the 1--5\,\mum\ photometry places a limit on the fractional number density of CO$_2$ of $\log(n_{\rm CO_2})\!<\!-2.7$, but provides no other constraints so far. We conclude that \roxsBb\ has mid-IR photometric features that are systematically different from other previously observed planetary-mass and field objects of similar temperature. It remains unclear whether this is in the range of the natural diversity of targets at the very young ($\sim$2\,Myr) age of \roxsBb\ or unique to its early evolution and environment.
}


\maketitle

\section{Introduction}
In the past decade, direct imaging observations of nearby young stars have revealed over a dozen companions with estimated masses between 4 and 13 \MJup\ \citep[][and references therein]{bow16}.    Multiwavelength photometry and/or spectroscopy of the first of these planetary-mass companions (e.g., HR 8799 bcde, 2M 1207 B; \citealt{mar08,mar10,cha04}) revealed key differences with field brown dwarfs, with color-magnitude diagram positions following a reddened extension of the L dwarf sequence to fainter magnitudes and cooler temperatures.  Compared to field objects of the same effective temperatures (early T dwarfs), their atmospheres are substantially cloudier and exhibit stronger non-equilibrium carbon chemistry \citep{cur11,gal11}.  The companions' low surface gravities drive these two effects, allowing clouds to extend higher up into the atmosphere and enhancing vertical mixing of carbon monoxide.  More generally, these objects and other similar ones reveal a gravity/age dependent transition from L type (cloudy and methane poor) to T type (mostly clear and methane rich) \citep{mar12}.  

More recently, analysis of hotter directly imaged planetary companions ($T_{\rm eff}$ $\sim$ 1500--2500 $K$) like $\beta$ Pictoris b and other young L-type objects have also revealed differences with old field L dwarfs.  Typically, these objects lie redward of the field sequence in both near- and mid-infrared color-magnitude diagrams \citep{cur13,fah16} and are likewise thought to be particularly cloudy and -- like the HR 8799 planets -- low gravity. Analyzing other directly imaged L-type planetary companions allows us to determine whether they share these characteristics.

While  directly imaged planets have typically been analyzed using atmosphere forward modeling \citep[e.g.,][]{cur11}, some recent studies have instead adopted a ``spectral retrieval'' method.  With retrieval methods, the key features of a planet's atmosphere -- surface gravity, composition, and pressure-temperature profile -- are treated as free parameters.  Although the input physics is simpler than that of forward models, retrieval allows robust determinations of the uncertainty in individual parameters, such as the abundances of the main molecular constituents, the shape of the pressure-temperature profile, and the surface gravity. Retrieval has been applied toward modeling infrared photometry of transiting planets \citep[e.g.,][]{mad09,lin12,bar17} and has now been utilized for the directly imaged planets HR\,8799\,b and $\kappa$\,And\,b \citep{lee13,lav16,tod16}, both of which are at least 30\,Myr old but cover a wide range of temperatures \citep[800--2000\,K;][]{mar08,cur11,car13,hin13}.  

The wide-separation companion to \roxsB\ \citep[\roxsBb,][]{cur14a} provides another insight into the properties of the youngest and highest temperature planetary-mass objects.   First identified as a candidate companion by \citet{rat05} and subsequently confirmed to be a bound companion \citep{cur14a,kra14}, \roxsBb\ orbits a close binary M star \citep{sim95} that is a likely member of the 2--3\,Myr-old $\rho$~Ophiuchus star-forming cloud.  Given this age, adopting a distance of 135\,pc \citep{mam08}, and assuming standard planet cooling models \citep[e.g.,][]{bar03}, \roxsBb\ is most likely a $\sim$\,9$\pm$3\,\MJup\ companion orbiting at $\sim$\,150\,au, slightly more massive and at a somewhat wider orbital separation than HR\,8799\,b \citep[70\,au,][]{mar08}. Early analysis of \roxsBb's near-infrared colors reveals that it too is redder than the field L dwarf sequence \citep{cur14a}; its spectrum shows evidence for a low surface gravity \citep{cur14a,bow14}.   The first atmospheric modeling of \roxsBb\ argues in favor of thicker clouds than field L dwarfs of the same temperature and implies a mass broadly consistent with cooling model estimates \citep{cur14b}.

New thermal infrared photometry of \roxsBb\ expand the wavelength coverage for this planetary object and allow us to better probe trends found in other young planets. For example, in the case of $\beta$\,Pic\,b, atmospheric forward models fit to the planet's rather flat spectral energy distribution in the thermal infrared provided evidence not only of  thick clouds, but also of small dust grains in its atmosphere \citep{cur13}.  For the HR 8799 planets, 3--5\,\mum\ imaging probed non-equilibrium carbon chemistry \citep{gal11}.  Thermal infrared imaging of GSC\,06214\,B identified extremely red colors consistent with a circum(sub)stellar disk, corroborating evidence for an accretion disk from spectroscopy \citep{bow11}.  While somewhat shallow \Lp\ imaging of  \roxsBb\ has already been published \citep{kra14,cur14b}, deeper \Lp\ data and photometry obtained at longer wavelengths clarify the object's overall spectral energy distribution shape and may better point to trends in young hot planet atmospheres.   Analyzing \roxsBb\ with spectral retrieval methods may place  further  quantitative constraints on its molecular abundances.

Here we present new photometry in the \Lp\ ($\lambda_\mathrm{c}$=3.78\,\mum), \Bralpha\ ($\lambda_\mathrm{c}$=4.05\,\mum), and \Ms\ filters ($\lambda_\mathrm{c}$=4.67\,\mum) observed with the NIRC2 instrument on Keck. Section~\ref{sec:obs} describes the observations and data; Sect.~\ref{sec:analysis} presents our photometry as well as an analysis using spectral fitting and retrieval methods. Results are presented in Sect.~\ref{sec:results} and discussed in Sect.~\ref{sec:discussion}.

\section{Observations and data reduction}\label{sec:obs}
Photometry was obtained on June 8, 2015, using the NIRC2 infrared imaging instrument on the Keck telescope. The \emph{narrow} camera was used in pupil-stabilized mode featuring a pixel scale of 9.952$\pm$0.050\,mas/pix \citep{yel10}. Individual exposure times were 0.2, 1.0, and 0.1 seconds in \Lp, \Bralpha, and \Ms,\ respectively; the number of coadds were adjusted to add up to integration times of 20 (\Lp) and 30 seconds (\Bralpha, \Ms) per exposure and filter. Overall, we obtained total on-source integration times of 13$\times$20\,sec=260\,sec, 27$\times$30\,sec = 810\,sec, and 36$\times$30\,sec = 1080\,sec in \Lp, \Bralpha, and \Ms, respectively. A summary of the observational parameters is presented in Table~\ref{tab:obslog}. 

\begin{table*}
  \caption{Observing Log: 2015-06-15, Keck/NIRC2\label{tab:obslog}}
  \centering
  \begin{tabular}{lcllccccccc}
    \hline\hline
        {Object}&{Obs. Time}&{Filter}&{$t_\mathrm{int}$}&{$N_\mathrm{images}$}&{PA (start)} & {PA (end)}\\
        {} & {UT}&{} &{(s)} & & {($^{\circ}$)}& {($^{\circ}$)} \\
    \hline
ROXs 42B   & 07:20:17 & \Lp      & 20    & 13 & $-$42.99 & $-$38.96 \\
ROXs 42B   & 07:44:55 & \Ms      & 30    & 18 & $-$36.79 & $-$32.99 \\
ROXs 42B   & 08:03:34 & \Bralpha & 30    & 9  & $-$32.16 & $-$30.82 \\
HD 141569A & 08:16:18 & \Lp      & 50    & 74 & $-$27.16 & 18.75    \\
HD 141569A & 09:43:39 & \Lp      & 0.075 & 6  & \dots    & \dots    \\
HD 141569A & 09:46:11 & \Bralpha & 1     & 6  & \dots    & \dots    \\
HD 141569A & 09:48:42 & \Ms      &  0.1  & 6  & \dots    & \dots    \\
ROXs 42B   & 09:53:33 & \Ms      &  30   & 18 &   1.87   & 7.10     \\
ROXs 42B   & 10:11:16 & \Bralpha &  30   & 18 &   7.81   & 11.74    \\
  \hline
  \end{tabular}
\end{table*}

\begin{figure*}
\centering
\includegraphics[width=\textwidth]{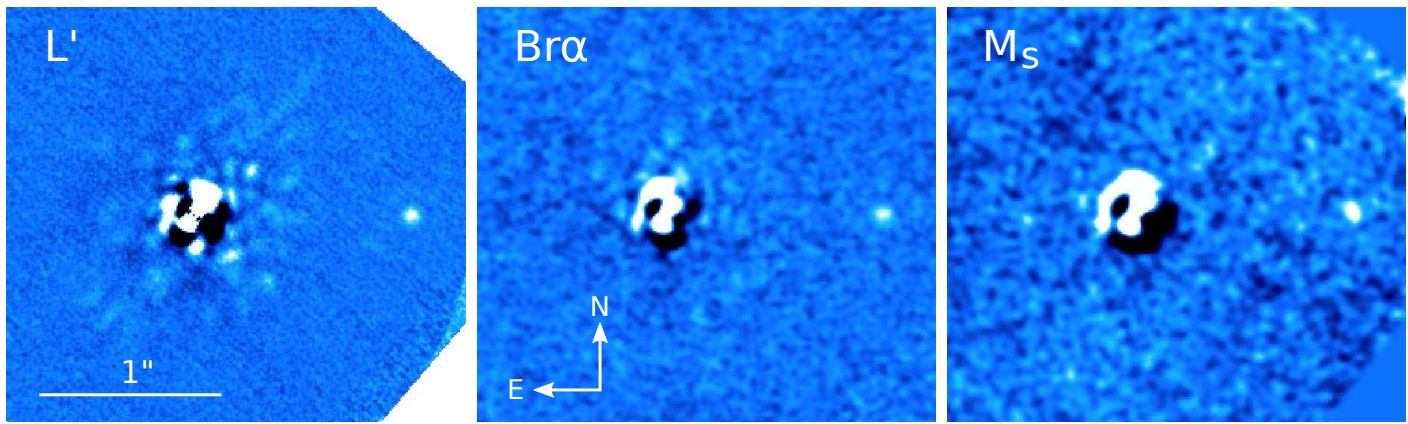}
\caption{New imaging observations of \roxsB\ and its companion in \Lp, \Bralpha, and \Ms\ filters. For illustration, a radially averaged profile was subtracted from the primary to reduce the seeing halo and the \Bralpha\ and \Ms\ images were smoothed with a 3-pixel  Gaussian. The color stretch is linear and cuts were adjusted to best display \roxsBb. North is up and east to the left.}
\label{fig:ImageStamps}
\end{figure*}

Individual images were bad-pixel corrected, flat-fielded, and sky subtracted. Sky frames were composed from on-source science exposures, which were taken in a 3-point dither pattern, through median combination. A distortion correction was applied using the \emph{nirc2dewarp.pro} \emph{IDL} routine\footnote{\url{http://www2.keck.hawaii.edu/inst/nirc2/dewarp.html}}. Modest field rotations of $\lesssim$$5^\circ$ were achieved in all filters. Accordingly, no high-angular resolution postprocessing algorithms (e.g., LOCI, PCA) were applied, but instead we derotated and stacked all images obtained in a filter to increase the signal-to-noise ratio (S/N). 
Figure~\ref{fig:ImageStamps} shows our final reduced images in the \Lp, \Bralpha, and \Ms\ filters after subtracting a radially averaged profile from the primary to reduce the seeing halo.

\section{Analysis}\label{sec:analysis}
\subsection{Photometry}
For photometric calibration, we observed HD\,141569 close in time. We compensated for the difference in airmass (target am$\approx$1.4--1.6, std am$\approx$1.1) using the corrections described in \citet{Tokunaga2002}, subtracting 0.05, 0.075, and 0.1\,mag from the calculated apparent \Lp, \Bralpha, and \Ms\ magnitudes, respectively. For this B9.5 spectral type star \citep{mer04} we assume intrinsic $H$$-$\Ms\,$\approx$\,0.0 and $H$$-$\Bralpha\,$\approx$\,0.0 colors (we note that both $J$$-$$H$ and $H$$-$\Ks\ are $<$0.05\,mag) and thus infer \Ms\,=\,\Bralpha\,=\,$H$\,=\,$6.86$\,mag. 

It was discovered during the analysis that the photometric standard shows \Bralpha\ in emission (Sean Brittain, private communication). While this has no strong effect on our \Lp\ photometry (\Bralpha\ emission line width $\sim$0.01\,\mum\ $\ll$ filter width $\approx$0.75\,\mum), we cannot use HD\,141569 to calibrate our 4.05\,\mum\ photometry. Instead, we measure the magnitude difference between \roxsBb\ and its host star using PSF photometry with the primary as PSF reference. The measured $\Delta$\Bralpha=5.53$\pm$0.05\,mag is translated into an apparent magnitude of \Bralpha=13.90$\pm$0.08\,mag by estimating the primary's \Bralpha\ brightness from its \Ks-band magnitude and \Ks$-$\Bralpha\ color. The latter we determine as \Ks$-$\Bralpha=0.298{\,$\pm$\,0.06\,mag} by linearly interpolating the expected \Ks$-$W1=0.11$\pm$0.01\,mag and \Ks$-$W2=0.17$\pm$0.01\,mag colors of a young M0-type star \citep{pec13} at 4.05\mum\ [(\Ks$-$\Bralpha)$^{\rm intrinsic}$=0.159\,mag] and adding a reddening of E(\Ks$-$\Bralpha)=0.139\,mag using an $R_\mathrm{V}$=3.1 extinction law \citep[]{mat90} and the \roxsB\ extinction of $A_\mathrm{V}$\,=\,1.9$^{+0.3}_{-0.2}$\,mag \citep{cur14a}. The uncertainty of the \Ks$-$\Bralpha\ color is calculated from the uncertainty of both the interpolation and the dereddening.

We derive dereddened absolute magnitudes of \roxsBb\ assuming a distance of 135\,$\pm$5\,pc from \citet{mam08} and assuming that the extinction is the same as for the primary. Table~\ref{tab:photometry} shows our new $>$3\,\mum\ photometry together with previous observations at shorter wavelengths. The new \Lp\ photometry is consistent with previous measurements by \citet{cur14b}. 

\begin{table*}
\caption{Photometry of \roxsB\ and \roxsBb\label{tab:photometry}}
\centering
\begin{tabular}{lccr@{\,$\pm$\,}lr@{\,$\pm$\,}lr@{\,$\pm$\,}lr@{\,$\pm$\,}lcc}
  \hline\hline
&
  &
  &
  \multicolumn{2}{c}{primary}&
  \multicolumn{2}{c}{Companion}&
  \multicolumn{2}{c}{apparent}&
  \multicolumn{2}{c}{Abs Mag} &
  &
  \\
  UT Date &
  Telescope/Camera &
  Filter &
  \multicolumn{2}{c}{mag} &
  \multicolumn{2}{c}{$\Delta$mag} &
  \multicolumn{2}{c}{Mag} &
  \multicolumn{2}{c}{(dereddened)} &
  S/N\tablefootmark{a}&
  ref. \\
\hline
2011-Jun-22 & Keck/NIRC2 & $J$       & 9.906 & 0.020\tablefootmark{b} &         7.00 & 0.11         & 16.91 & 0.11 &10.75 & 0.17 & 11$^{c}$  & 2  \\  
2011-Jun-22 & Keck/NIRC2 & $H$       & 9.017 & 0.020\tablefootmark{b} &         6.86 & 0.05         & 15.88 & 0.05 & 9.87 & 0.14 & 33$^{c}$  & 3  \\  
2011-Jun-22 & Keck/NIRC2 & \Ks       & 8.671 & 0.020\tablefootmark{b} &         6.34 & 0.06         & 15.01 & 0.06 & 9.14 & 0.14 & 52$^{c}$  & 2  \\  
2016-Jun-08 & Keck/NIRC2 & \Lp       & 8.42  & 0.05                   &         5.55 & 0.06         & 13.97 & 0.06 & 8.23 & 0.10 & 22.2      & 1  \\  
2016-Jun-08 & Keck/NIRC2 & \Bralpha  & 8.37  & 0.06                   &         5.53 & 0.05         & 13.90 & 0.08 & 8.18 & 0.12 & 14.5      & 1  \\  
2016-Jun-08 & Keck/NIRC2 & \Ms       & \multicolumn{2}{c}{\dots  }    & \multicolumn{2}{c}{\dots  } & 14.01 & 0.23 & 8.31 & 0.24 &  6        & 1      \\
\hline
\end{tabular}
\tablefoot{
  \tablefoottext{a}{Signal-to-noise ratio as derived in Sect.~\ref{sec:SN}}
  \tablefoottext{b}{From the 2MASS catalog \citep{cut03}}
  \tablefoottext{c}{Signal-to-noise ratio for detections published in \citet{cur14a,cur14b}}
\tablebib{
(1) This paper; (2) \citealt{cur14b}; (3) \citealt{cur14a}
}}
\end{table*}

\subsection{Sensitivity and signal-to-noise ratio\label{sec:SN}}
To estimate the S/N, we measure the flux in 16 apertures\footnote{In the case of \Lp,\ three apertures were too close to the edge of the detector and have been disregarded} of radius 0\farcs043 (\Lp, \Bralpha) and 0\farcs060 (\Ms), distributed along a circle of 0\farcs28--0\farcs35 radius centered on the companion. Comparing the standard deviation of the noise measurement to the flux in an aperture of the same size centered on the companion, we infer S/N of 22.2, 14.5, and 5.7 for our detections of \roxsBb\ in \Lp, \Bralpha, and \Ms, respectively.

\begin{figure}
\centering
\includegraphics[width=0.9\columnwidth]{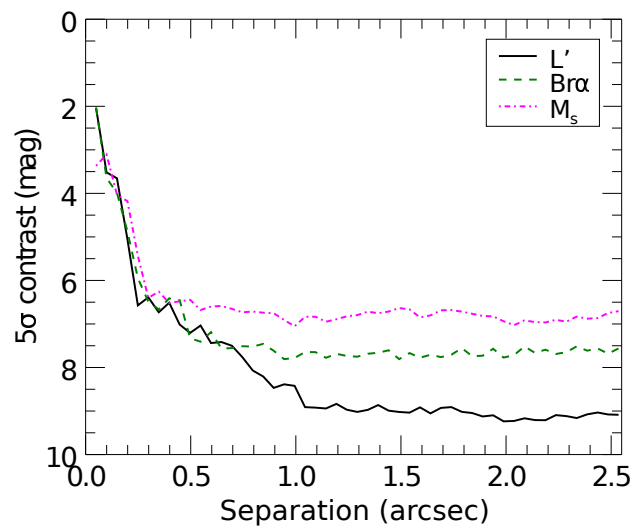}
\caption{Five sigma contrast curves as described in Sect.~\ref{sec:SN}.}
\label{fig:sensitivity}
\end{figure}

Fig.~\ref{fig:sensitivity} shows our 5$\sigma$ sensitivity to the detection of any additional companions in the field of view. 
Contrast was derived as in  the above scheme, but noise apertures were distributed on circles with radii between 0\farcs05 and 2\farcs5 around the primary star and the number of apertures was set to the maximum possible number fitting on this circle with no overlap between neighboring apertures. The values are listed in Table~\ref{tab:sensitivity}.
\begin{table}
\caption{5$\sigma$ detection limits\label{tab:sensitivity}}
  \centering
  \begin{tabular}{lcccccc}
\hline\hline
  &
\multicolumn{6}{c}{$\Delta$ (mag) at $\rho$ =}
\\
{Filter} &
{0\farcs10}  &
{0\farcs25}  &
{0\farcs50}  &
{0\farcs75}  &
{1\arcsec}   &
{2\arcsec}  \\ 
\hline
\Lp              &  3.5 &  6.6 &  7.2 &  7.8 &  8.4 &  9.2 \\
\Bralpha         &  3.7 &  6.0 &  7.4 &  7.5 &  7.7 &  7.7 \\
\Ms              &  3.1 &  5.5 &  6.6 &  6.7 &  7.1 &  7.0 \\
\hline
  \end{tabular}
\end{table}

\subsection{Model fitting}\label{sec:models}
We compare the photometric SED of \roxsB, including the new mid-IR photometry, as well as published K-band (1.95--2.45\,\mum) spectroscopy \citep{cur14a} to stellar atmosphere models. In order to explore whether any additional constraints can be obtained from the new mid-infrared photometry, we select a similar set of models to those used in \citet{cur14b}. We compare three types of models, those with dust-free atmospheres (AMES-COND, BT-COND), models with uniformly dusty atmospheres (AMES-DUSTY, BT-DUSTY), and models with a cloud model (for references see Table~\ref{tab:models}). The third category includes BT-SETTL and our own models, hereafter called Burrows models, which feature a thick cloud layer and small dust grains. Where possible, models were retrieved from the \emph{Theoretical spectra server}\footnote{\url{svo2.cab.inta-csic.es/theory/newov/}} for consistent data formats (all except BT-SETTL~(2015) and the Burrows models), spanning effective temperatures of \teff=1500\,K\dots3000\,K in steps of 100\,K, surface gravities of \logg=2.5\dots5.5 in steps of 0.5, and keeping metallicity fixed at a solar level. Other metallicity values were not available for most of the models, in particular none in the interesting temperature range of 1500\,K--2500\,K. Spectra from the BT-SETTL~(2015) grid were retrieved directly from the respective web repositories\footnote{\url{https://phoenix.ens-lyon.fr/Grids/BT-Settl/}} covering the same parameter range. Burrows models span a smaller range from \logg=3.4 to 4 and \teff=1900\dots2400, spaced by 50\,K ($<$2000\,K) and 100\,K ($>$2000\,K). See Table~\ref{tab:models} for a summary of the models and probed parameter space.

For every parameter combination (\teff,\logg) we calculate $\chi^2=\sum_i w_i((f_{\mathrm{data},i}-c\cdot f_{\mathrm{model},i})/\sigma_i)^2$, where $f_{\mathrm{data},i}$ are the photometric data from Table~\ref{tab:photometry}, $\sigma_i$ are the measurement uncertainties, and $c=(R/10\,\mathrm{pc})^2$ is a dilution factor dependent on the radius $R$ of the object. In the case of photometry, $f_{\mathrm{model},i}$ is obtained by multiplying the model spectrum with the respective filter curves. To compare the models with spectroscopy, we use the $K$-band spectrum of \roxsBb\ as presented by \citet{cur14a}, i.e., binned to the resolution of the Burrows models of $R\approx300$. Model spectra of the other families were rebinned to the same resolution. To account for the varying number of resolution elements in each filter, every data point is weighted with $w_i$ proportional to the FWHM of the photometric filter or the width of the spectroscopic resolution element, respectively \citep{cus08}. The validity of $\chi^2$ is conserved by normalizing the weighting factors so that $\sum w_i=N$, identical to the number of data points. In the case of fitting the combined spectroscopy and photometry, weights for the photometric filters are $w_J$=18, $w_H$=33, $w_{K_\mathrm{s}}$=34, $w_{L^\prime}$=77, $w_{\mathrm{Br}\alpha}$=7.2, and $w_{M_\mathrm{s}}$=28, while a spectroscopic element is weighted with $w_\mathrm{spec}$=0.18. Because the spectroscopy was calibrated with the $K$-band photometry, it entered the combined fit as $\chi^2=\sum_i w_i((f_{\mathrm{phot},i}-c\cdot f_{\mathrm{model},i})/\sigma_i)^2 + \sum_j w_{\rm spec}((f_{\mathrm{spec},j}-c^*\cdot f_{\mathrm{model},j})/\sigma_j)^2$ for $i=\{1,\dots,6\}$ photometric and $j=\{1,\dots,292\}$ spectroscopic data points. The scaling factor $c^*$ minimizes the second summand and may be different from $c$. This way we avoid giving the $K$-band photometry overproportional weight as it has already been used for absolute calibration of the spectrum. 
  
Prior to fitting, our photometry and spectroscopy were dereddened assuming $A_\mathrm{V}$=1.9\,mag and an $R_\mathrm{V}$=3.1 extinction law \citep[]{mat90} and converted to flux density units using the zero points in \citet{cur13}. We note that assuming an extinction of $A_\mathrm{V}$=1.7\,mag as derived by \citet{bow14} does not lead to significantly different fitting results. The K-band spectroscopy was flux calibrated by scaling it so that integration over the 2MASS \Ks\ filter curve reproduces the \roxsBb\ \Ks\ magnitude (Table~\ref{tab:photometry}). The uncertainty of the calibration enters our fit as an uncertainty of the derived radius ($\sim$15\%).

Best fits for every model in the grid were determined through $\chi^2$ minimization using the \emph{mpfit.pro} procedure in \emph{IDL} \citep{mar09} with the dilution factor $c$ as the parameter of interest. This implies $n_\mathrm{dof}$=$N$$-$1 degrees of freedom (where $N$ can assume either the number of photometric data points,  the number of spectral resolution elements, or their sum,  depending on the statistical test we perform), which is used to calculate the statistical significance $P$ for each best-fit model, i.e., the chance to obtain the measured $\chi^2_{\rm min}$ assuming that the model represents the data. For $P<0.05$, the hypothesis that the model is an acceptable representation of the data is rejected with 95\% confidence.

\begin{figure*}
\centering
\setlength{\unitlength}{\textwidth}
\begin{picture}(1,0.68)
  \put(0.00,0.35){\includegraphics[width=0.49\textwidth]{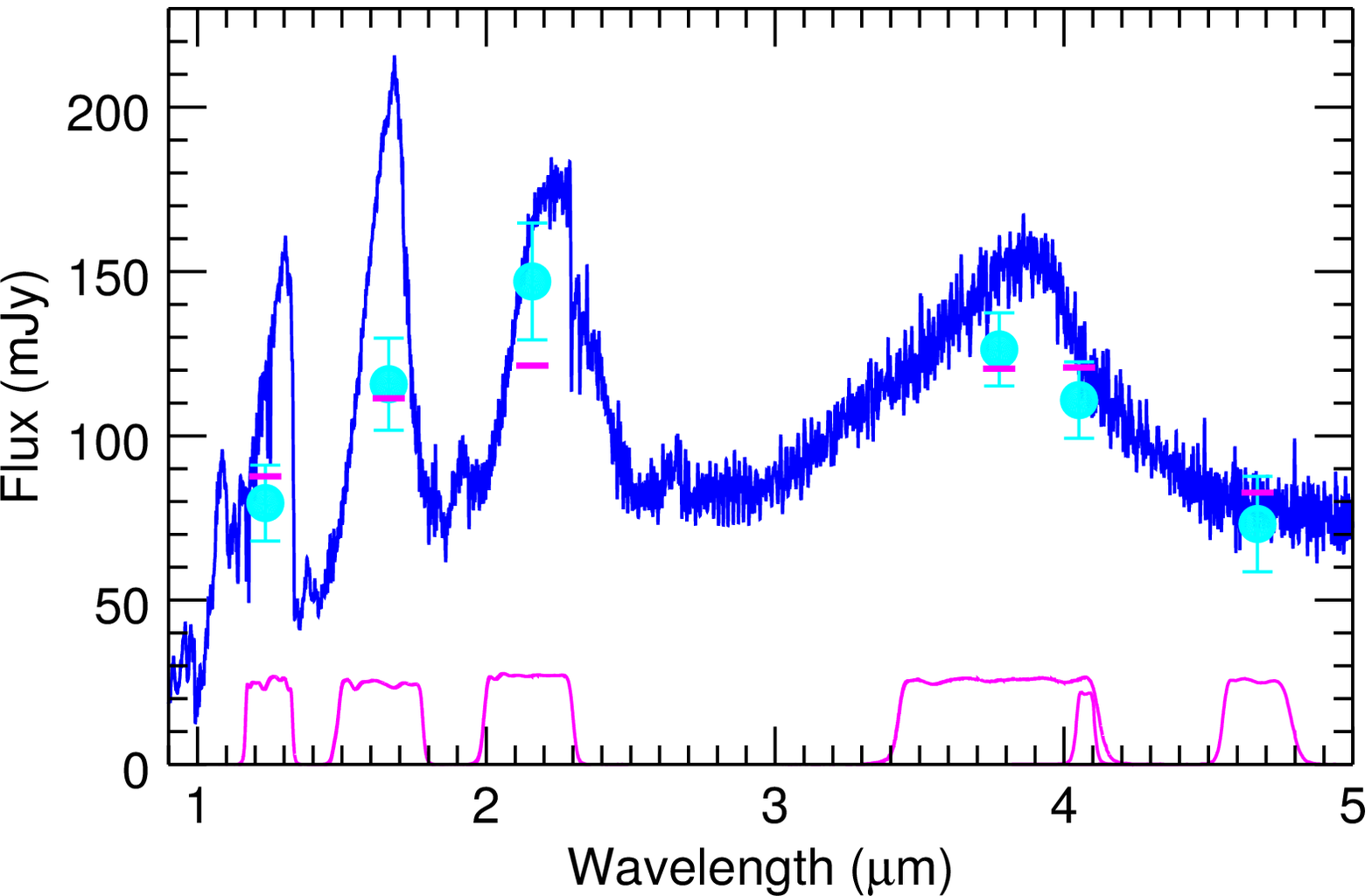}}
  \put(0.09,0.63){\fontfamily{phv}\selectfont AMES-DUSTY (Best fit: \teff=2000\,K, \logg=5.0)}
  \put(0.52,0.35){\includegraphics[width=0.49\textwidth]{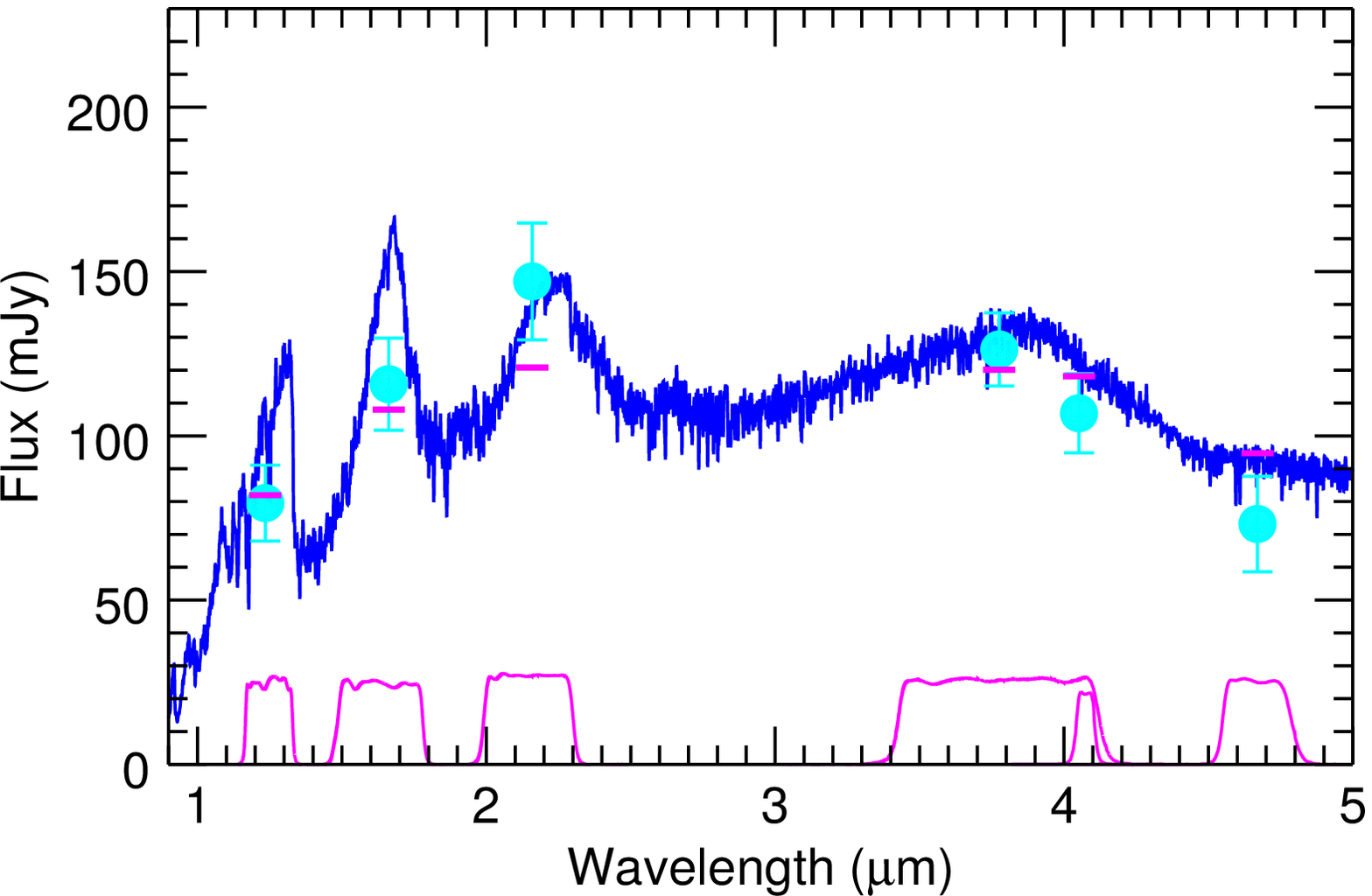}}
  \put(0.61,0.63){\fontfamily{phv}\selectfont BT-DUSTY (Best fit: \teff=1900\,K, \logg=4.5)}
  \put(0.00,0.00){\includegraphics[width=0.49\textwidth]{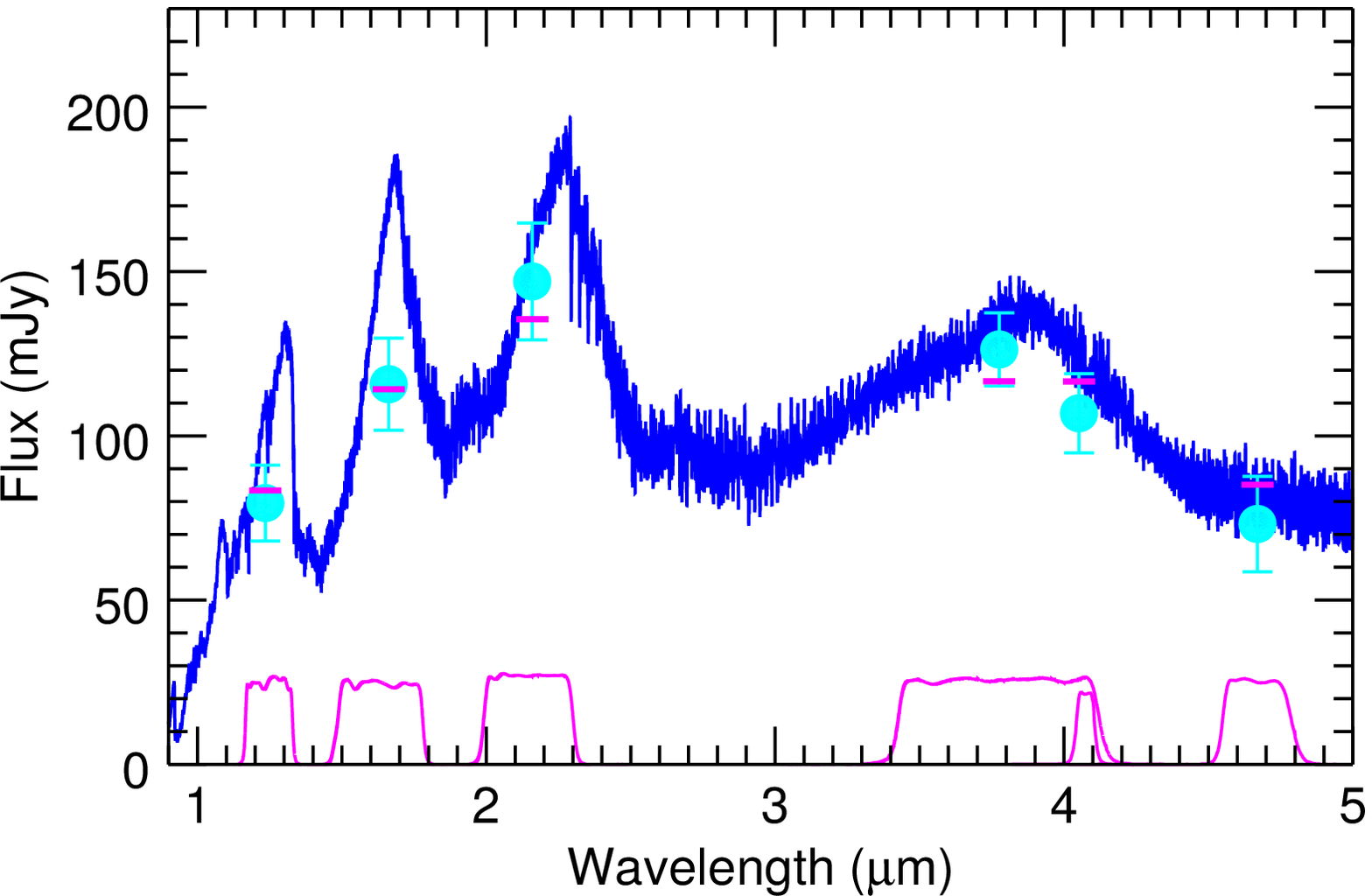}}
  \put(0.09,0.28){\fontfamily{phv}\selectfont BT-SETTL (Best fit: \teff=2000\,K, \logg=3.0)}
  \put(0.52,0.00){\includegraphics[width=0.49\textwidth]{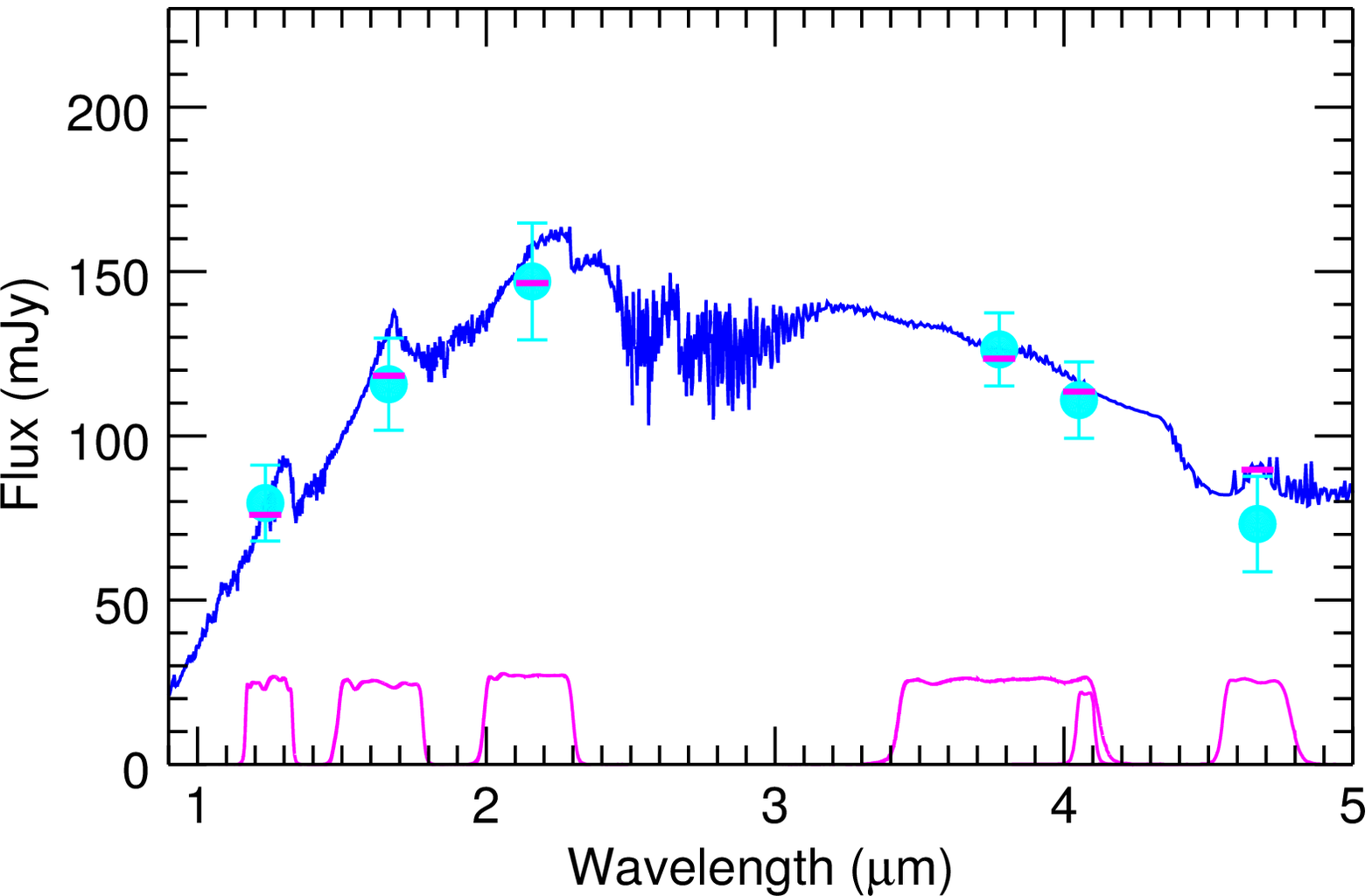}}
  \put(0.61,0.28){\fontfamily{phv}\selectfont Burrows (Best fit: \teff=1900\,K, \logg=3.4)}
\end{picture}
\caption{Best-fit SEDs (blue), data (cyan), model photometry (magenta). Effective temperature and \logg\ of each best fit are indicated in each panel. Filter transmission curves are at the bottom of each plot.}
\label{fig:spectralfit}
\end{figure*}

\begin{figure*}
\centering
\setlength{\unitlength}{\textwidth}
\begin{picture}(1,0.65)
  \put(0.00,0.0){\includegraphics[width=1\textwidth]{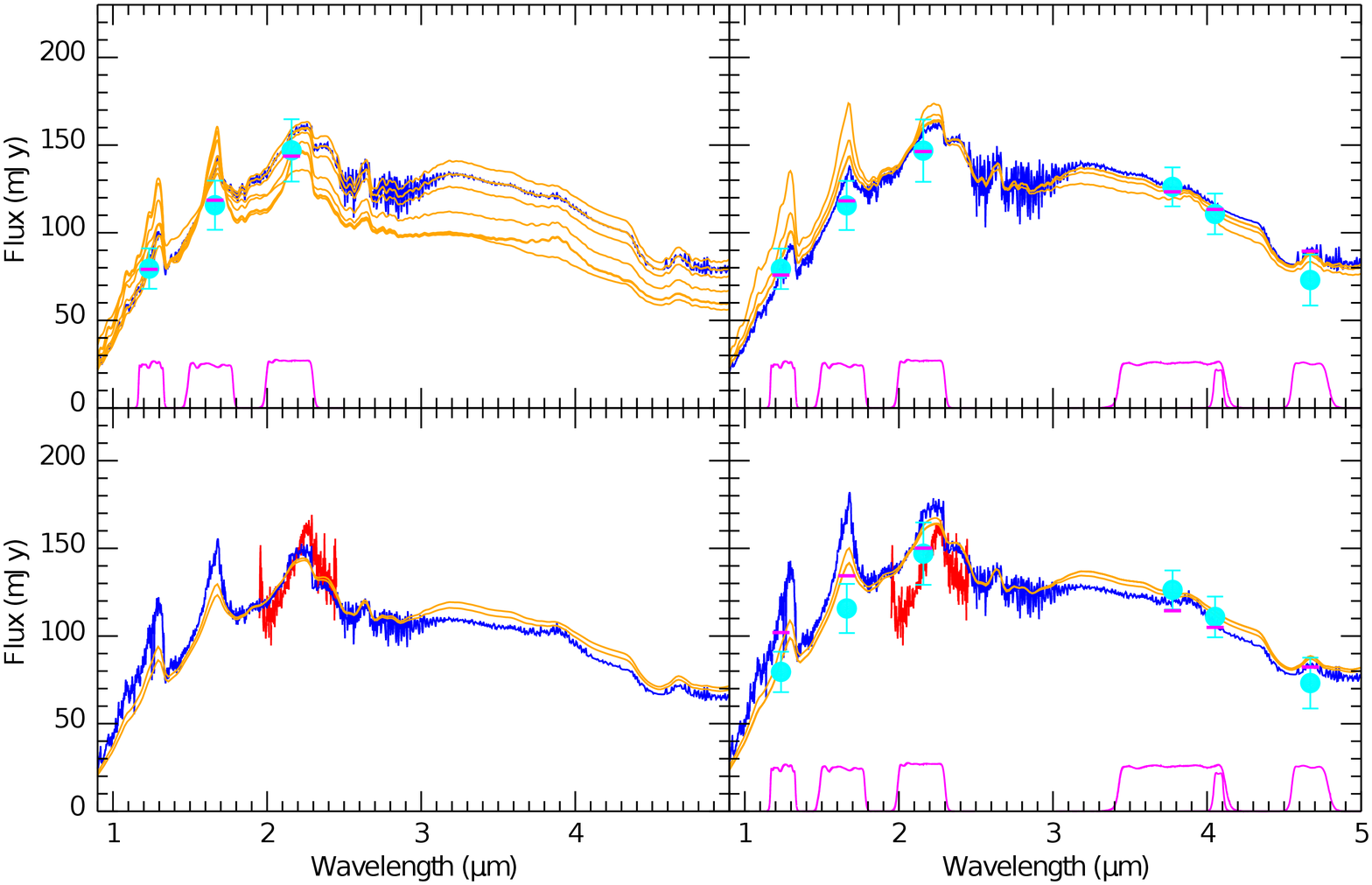}}
  \put(0.105,0.60){\fontfamily{phv}\selectfont a) $JH$\Ks\ photometry}
  \put(0.57,0.60){\fontfamily{phv}\selectfont b) $JH$\Ks\Lp\Bralpha\Ms\ photometry}
  \put(0.105,0.305){\fontfamily{phv}\selectfont c) $K$-band spectroscopy}
  \put(0.57,0.305){\fontfamily{phv}\selectfont d) all photometry+spectroscopy}
\end{picture}
\caption{Best-fit Burrows model (blue spectrum) and all other Burrows models with $P>0.05$ (yellow, smoothed for easier visibility). {\em a)} Range of accepted models when only $JH$\Ks\ photometric data are fit. {\em b)} The mid-infrared spread of acceptable model fits is strongly reduced when all 1--5\,\mum\ photometric points are taken into account. {\em c)} Fits of the $K$-band spectroscopy (red). {\em d)} Simultaneous fit to all $JH$\Ks\Lp\Bralpha\Ms\ photometry and $K$-band spectroscopy.}
\label{fig:comparefilters}
\end{figure*}

The ranges of \teff\ and \logg\ of all models that are not rejected are listed in Table~\ref{tab:fittingresults}  together with the best-fit parameters for fits of the $JH$\Ks\Lp\Bralpha\Ms\ photometry, the K-band spectroscopy, and their combination, respectively. Fig.~\ref{fig:spectralfit} shows the best fits to all available $JH$\Ks\Lp\Bralpha\Ms\ photometry.
\begin{table}
  \caption{Used models and input parameter ranges\label{tab:models}}
  \centering
  \begin{tabular}{lccc}
\hline\hline
  Model family &
  \teff &
  \logg &
  Ref.\\
\hline
\multicolumn{4}{l}{\em Without dust in the atmosphere}\\
AMES-COND        & 1500--3000 & 2.5--5.5 & 1,2\\[0.5ex]
BT-COND          & 1500--2500 & 4.0--5.5 & 3  \\
                 & 2600--3000 & 2.5--5.5 &    \\[2ex]
\multicolumn{4}{l}{\em With atmospheric dust} \\
AMES-DUSTY       & 1500--3000 & 3.5--5.5 & 1,4\\[0.5ex]
BT-DUSTY         & 1500--2500 & 4.5--5.5 & 3  \\
                 & 2600--3000 & 2.5--5.5 &    \\[2ex]
\multicolumn{4}{l}{\em With a cloud model}    \\
BT-SETTL         & 1500--1900 & 3.5--5.5 & 3  \\
                 & 2000--3000 & 2.5--5.5 &    \\[0.5ex]
BT-SETTL-2015    & 1500--1900 & 3.5--5.5 & 5  \\
                 & 2000--3000 & 2.5--5.5 &    \\[0.5ex]
Burrows          & 1900--2400 & 3.4--4.0 & 6  \\
\hline
\end{tabular}
\tablebib{(1) \citealt{all01}; (2) \citealt{bar03}; (3) \citealt{all12}; (4) \citealt{cha00}; (5) \citealt{bar15}; (6) \citealt{cur13}.}
\end{table}
\begin{table*}
  \caption{Fitting results\label{tab:fittingresults}}
  \centering
  \begin{tabular}{l|ccccc|ccc}
\hline\hline
  &
  \multicolumn{5}{c}{best fit} &
  \multicolumn{3}{c}{range\tablefootmark{a}} \\
  Model family &
  \teff &
  \logg &
  $R$ ($R_\mathrm{Jup}$) &
  $\chi^2_\mathrm{min}$ &
  $P(\chi^2_\mathrm{min})$ &
  \teff &
  \logg &
  $R$ ($R_\mathrm{Jup}$) \\
\hline
\multicolumn{9}{c}{\em $J$$H$\Ks\Lp\Bralpha\Ms\ photometry ($N$=6)}\\[0.5ex]
AMES-COND       & 1700 & 2.5 & 3.4 &10.0 & 0.07  &1700              & \emph{2.5}             & 3.4       \\[0.5ex]
BT-COND         & 1800 & 4.0 & 3.1 &14.1 & 0.01  &   \dots          &   \dots                &  \dots    \\[0.5ex]
AMES-DUSTY      & 2000 & 5.0 & 2.5 & 4.1 & 0.54  &1800--2100        & \emph{3.5}--\emph{5.5} & 2.4--2.8  \\[0.5ex]
BT-DUSTY        & 1900 & 4.5 & 3.0 & 5.5 & 0.36  &1700--2000        & \emph{4.5}--\emph{5.5} & 2.6--3.2  \\[0.5ex]
BT-SETTL        & 2000 & 3.0 & 2.4 & 2.5 & 0.78  &1600              & 4.5                    & 3.4       \\
                &      &     &     &     &       &1800--2100        & \emph{2.5}--\emph{5.5} & 2.3--3.2  \\[0.5ex]
BT-SETTL-2015   & 1900 & 4.5 & 2.6 & 2.3 & 0.81  &1700--2100        & \emph{2.5}--\emph{5.5} & 2.4--3.1  \\[0.5ex]
Burrows         & 1900 & 3.4 & 2.2 & 1.5 & 0.91  &\emph{1900}--2100 & \emph{3.4}--\emph{4.0} & 2.0--2.7  \\[2ex]
\multicolumn{9}{c}{\em $K$-band spectroscopy ($N$=292)}  \\[0.5ex]
AMES-COND       & 2800 & 4.0 & 1.5 & 265 & 0.86  &2700--\emph{3000} & 3.5--4.5               & 1.3--1.6  \\[0.5ex]
BT-COND         & 2600 & 3.5 & 1.6 & 212 &  1    &2500--2800        & \emph{2.5}--4.5        & 1.4--1.7  \\[0.5ex]
AMES-DUSTY      & 2700 & 3.5 & 1.5 & 252 & 0.95  &1600--1700        & \emph{5.5}             & 2.9--3.2  \\
                &      &     &     &     &       &2600--2900        & \emph{3.5}--4.5        & 1.3--1.6  \\[0.5ex]
BT-DUSTY        & 2300 & 4.5 & 1.8 & 211 &  1    &1600--2000        & \emph{4.5}--5.0        & 2.6--4.7  \\
                &      &     &     &     &       &2300              & \emph{4.5}             & 1.8       \\
                &      &     &     &     &       &2600--2800        & \emph{2.5}--4.5        & 1.4--1.8  \\[0.5ex]
BT-SETTL        & 2400 & 2.5 & 1.8 & 173 &  1    &\emph{1500}--1800 & \emph{3.5}             & 2.9--4.1  \\
                &      &     &     &     &       &   1700           & \emph{3.5}--5.0        & 3.0--3.3  \\
                &      &     &     &     &       &2000--2800        & \emph{2.5}--4.5        & 1.4--2.3  \\[0.5ex]
BT-SETTL-2015   & 2700 & 3.5 & 1.5 & 234 &  1    &\emph{1500}--1600 & \emph{2.5}--3.0        & 3.7--4.6  \\
                &      &     &     &     &       &\emph{1500}--1600 & 4.0                    & 3.4--3.7  \\
                &      &     &     &     &       &2400--2900        & 3.0--4.5               & 1.3--1.8  \\[0.5ex]
Burrows         & 2100 & 3.6 & 1.9 & 316 & 0.15  &1950--2100        & \emph{3.4}--3.6        & 1.9       \\[2ex]
\multicolumn{9}{c}{\em Spectroscopy+photometry ($N$=298)}\\[0.5ex]
AMES-COND       & 2800 & 4.0 & 1.5 & 338 & 0.05  &2800              & 4.0                    & 1.5       \\[0.5ex]
BT-COND         & 2600 & 3.5 & 1.7 & 259 & 0.95  &2600--2700        & \emph{2.5}--4.5        & 1.5--1.7  \\[0.5ex]
AMES-DUSTY      & 2700 & 3.5 & 1.6 & 315 & 0.23  &2600--2800        & \emph{3.5}--4.0        & 1.5--1.7  \\[0.5ex]
BT-DUSTY        & 1800 & 4.5 & 3.4 & 234 &  1    &1600--2000        & \emph{4.5}--5.0        & 2.6--3.9  \\
                &      &     &     &     &       &   2300           & \emph{4.5}             & 1.9       \\
                &      &     &     &     &       &2600--2700        & \emph{2.5}--4.5        & 1.5--1.7  \\[0.5ex]
BT-SETTL        & 2300 & 2.5 & 2.0 & 196 &  1    &\emph{1500}--1800 & \emph{3.5}             & 2.8--3.6  \\
                &      &     &     &     &       &   1700           & \emph{3.5}--5.0        & 2.8--3.4  \\
                &      &     &     &     &       &2000--2700        & \emph{2.5}--4.5        & 1.5--2.4  \\[0.5ex]
BT-SETTL-2015   & 2600 & 3.5 & 1.6 & 294 & 0.54  &\emph{1500}       & 3.0                    & 3.8       \\
                &      &     &     &     &       &2500--2800        & 3.0--4.0               & 1.4--1.8  \\[0.5ex]
Burrows         & 2100 & 3.6 & 2.0 & 323 & 0.14  &1950--2100        & \emph{3.4}--3.8        & 2.0--2.5  \\
\hline
\end{tabular}
\tablefoot{
  \tablefoottext{a}{Parameter range of models that are consistent with the data according to their significance $P$$>$0.05 (see Sect.~\ref{sec:models}). Numbers in italics are at the limit of the probed model parameter space (see Table~\ref{tab:models}).}
}
\end{table*}
In Fig.~\ref{fig:comparefilters} we show all best-fit models for the example of the Burrows models to illustrate the diversity of spectra that satisfy $P>0.05$. The panels show all selected models for the following four cases: a) when only near-infrared photometry is fit, b) when using fits to all $JH$\Ks\Lp\Bralpha\Ms\ photometry, c) when fitting $K$-band spectroscopy alone, and d) when fitting all 1--5\,\mum\ photometry and $K$-band spectroscopy. It can clearly be observed how the choices of input data constrain the range of acceptable model fits.

\subsection{Retrieval}
\label{sec:retrieval}
To investigate how the mid-IR photometry constrains the atmospheric properties of \roxsBb, we adapted the spectral retrieval code described in \citet{tod16} to accept photometric input.  We refer to \citeauthor{tod16} for a detailed description of the method, and here give a brief summary  and describe the adaptions made.

In order to fit the photometric points with a model, we require a computationally cheap approach that takes a very limited number of free parameters as input and produces a model emission spectrum. The approach we take utilizes the plain parallel approximation for the  companion's atmosphere, containing 50 layers spaced equally in the log of pressure. All layers contain the same abundances of the chemical species we consider, at a temperature and pressure specified by a pressure-temperature (P-T) profile. The number densities as a fraction of the total number of CH$_{4}$, CO, CO$_{2}$, H$_{2}$O, C$_2$H$_2$, HCN, and NH$_3$ are free parameters. Additional free parameters include the surface gravity of the object and the temperatures of six layers spaced equally (in $\log(P)$) throughout the atmosphere. The temperatures corresponding to the other layers are interpolated to create a smoothly varying P-T profile. We experiment with sampling the pressure range by dividing it into more layers and  fewer layers, but we find no impact on the upper limits we are able to place on the number densities of molecules. For CO, CO$_{2}$, and H$_{2}$O we adopt the opacities from the HITEMP data base \citep{rot10}, while those for CH$_{4}$, C$_2$H$_2$, HCN, and NH$_3$ are adopted from \citet{fre14}. Collision-induced opacities for H$_2$-H$_2$ and H$_2$-He are also included \citep{bor01, bor02}.

We assume a cloud-free atmosphere. The effect of clouds would be to make the spectrum appear flatter \citep[see, e.g.,][]{lee13} because emission from below the cloud deck, i.e., at high temperatures, is being absorbed. The strength of this effect as a function of wavelength depends on the assumed cloud model, e.g., particle material, size distribution, and shape, as well as the horizontal extent of the clouds (patchy versus full coverage), and their vertical extent and particle column density. While degeneracies introduced by clouds, for example between molecular abundances and cloud presence, may be resolved by high signal-to-noise spectroscopy studies \citep[e.g., L-band spectroscopy, see Figure~1 in][]{lee13}, the current data set does not constrain enough parameters to justify introducing a complex cloud model.

The fitting approach we employ to match this simple emission model to the observed photometry is a Markov chain Monte Carlo (MCMC) setup, based on the Metropolis-Hastings algorithm using a Gibbs sampler and a Gaussian trial value probability distribution \citep{ford05, ford06}. Our specific implementation is discussed in detail in \citet[][]{tod12}.  We assume flat priors for all free parameters. We run the chain for $10^6$ iterations and {in our analysis we discard the initial 10\%, where chain convergence occurs, to avoid biasing our results.}

\section{Results}\label{sec:results}
\subsection{Infrared colors of ROXS 42B b}\label{sec:results_colors}
We compare the near- to mid-IR photometry of \roxsBb\ with the photometry of field dwarfs from the compilation by \citet{leg10}, spanning spectral types between M7 and T5, equivalent to temperatures between $\sim$2500\,K and 700\,K. In addition, we include the photometry of young directly imaged planetary mass objects around $\beta$\,Pic \citep{cur13}, HR\,8799 \citep{mar08,mar10,cur11}, $\kappa$\,And \citep{car13}, GSC\,06214 \citep{ire11,bai13}, and RXS\,1609 \citep{laf10} using the photometry compiled by \citet{cur13}, and the updated HR\,8799 near-IR photometry by \citet{zur16}.

\begin{figure*}
\centering
\includegraphics[height=0.22\textheight]{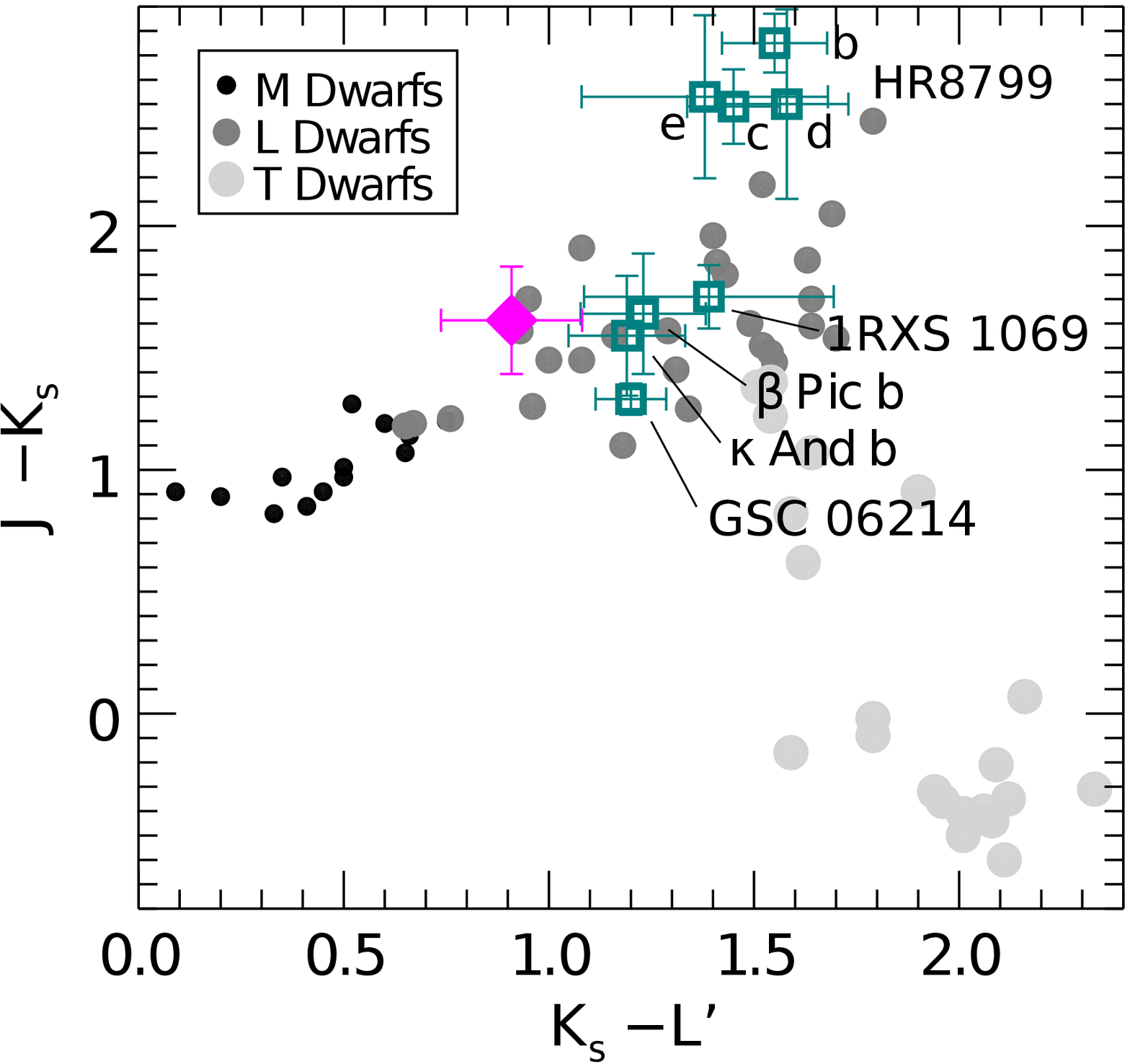}\hfill
\includegraphics[height=0.22\textheight]{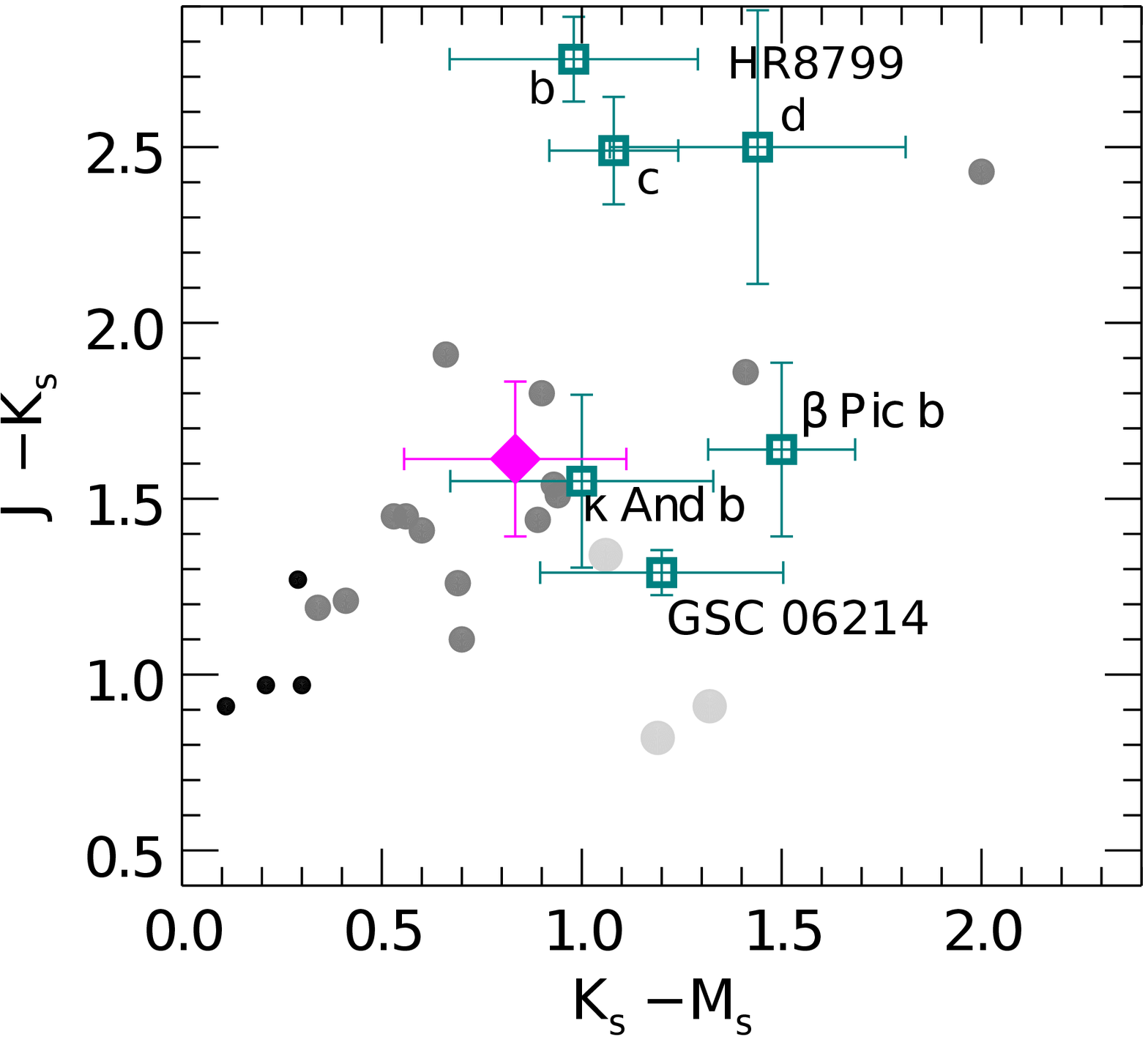}\hfill
\includegraphics[height=0.22\textheight]{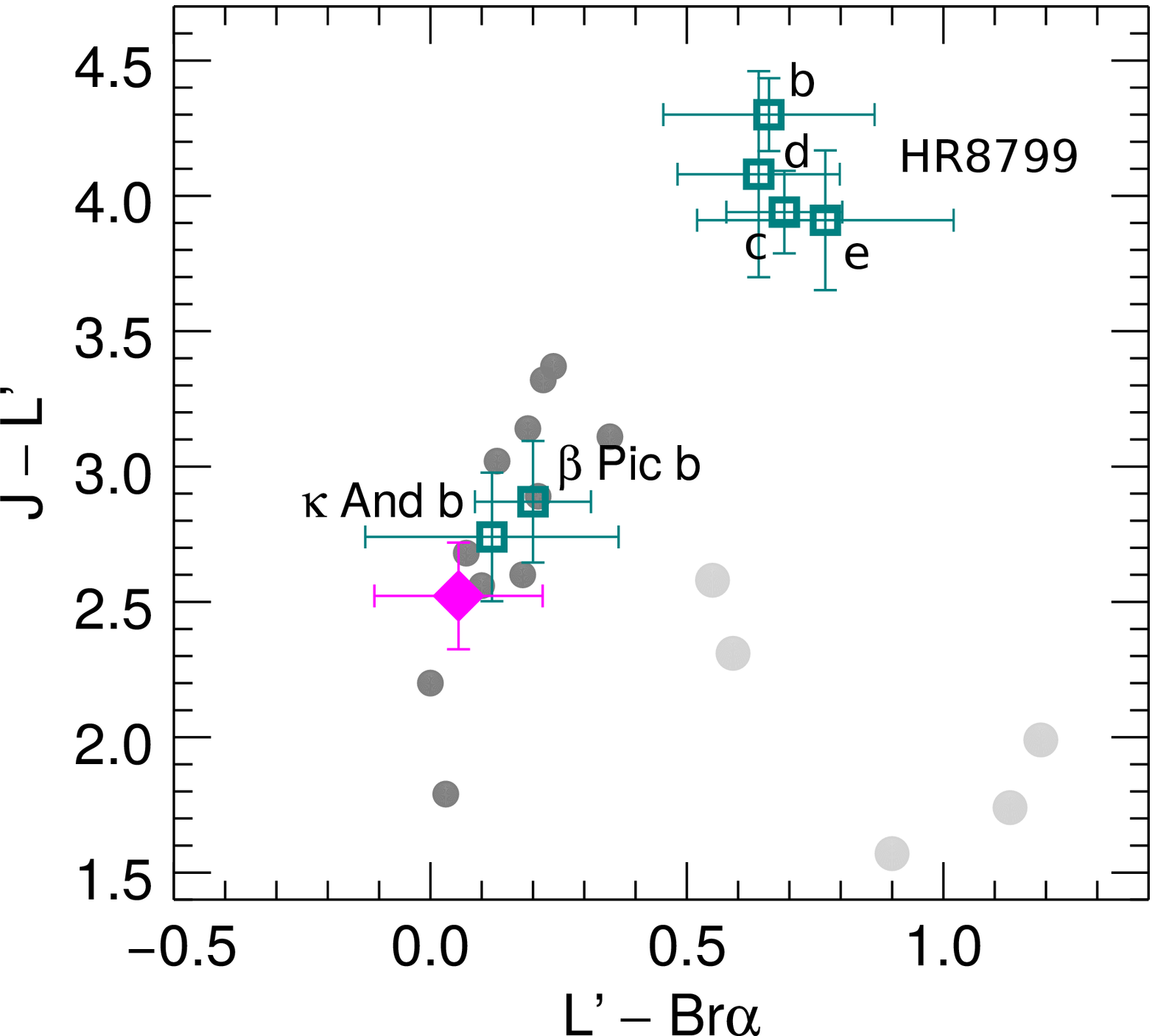}
\caption{Color-color diagrams comparing \roxsBb\ colors (magenta diamond) to those of field M, L, and T dwarfs (gray circles) from \citet{leg10}.  We also overplot the positions of young substellar objects/other directly imaged planets \citep[turquoise squares,][]{cur13}.}
\label{fig:colorcolor}
\end{figure*}

\begin{figure*}
\centering
\includegraphics[height=0.22\textheight]{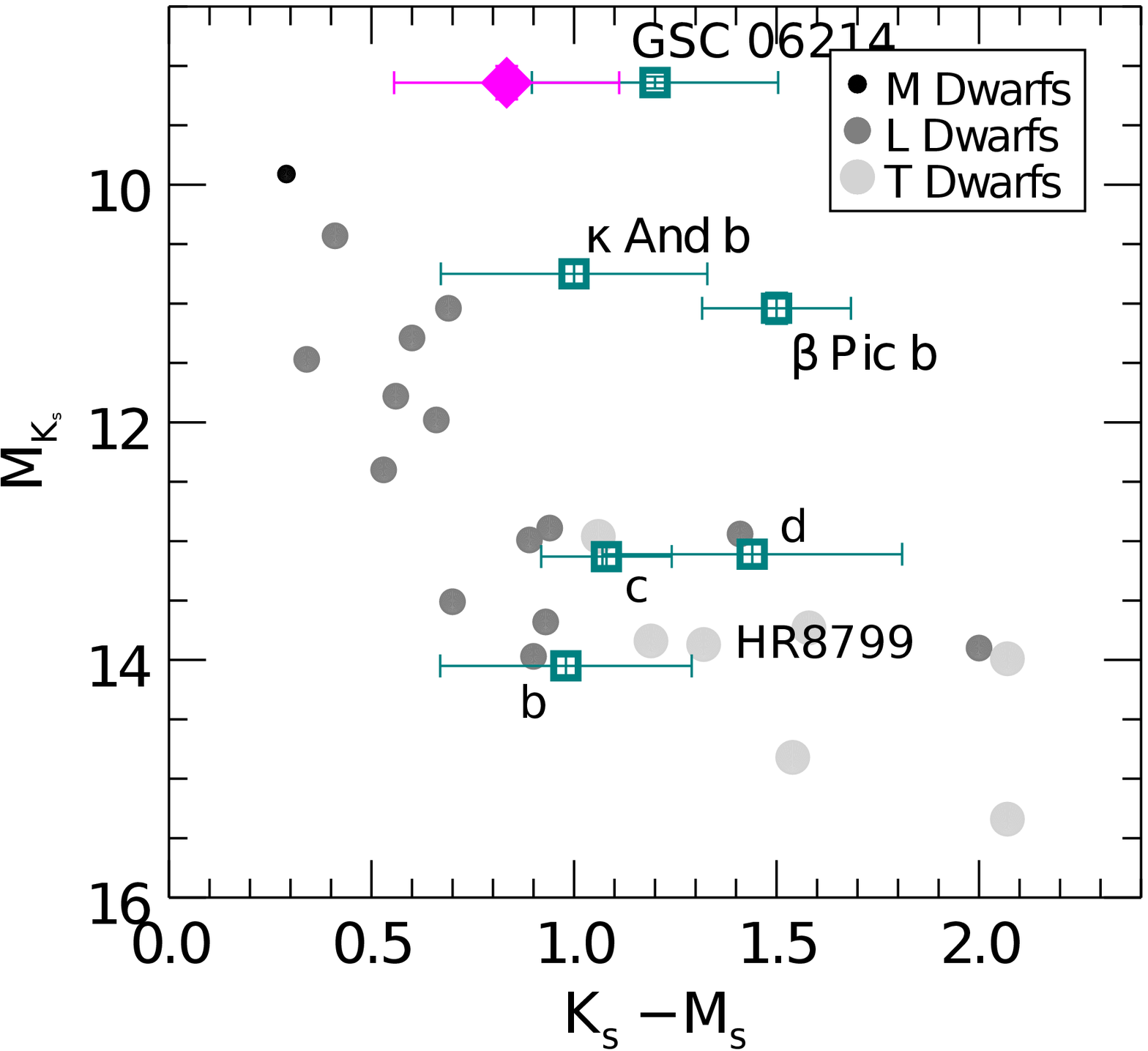}\hfill
\includegraphics[height=0.22\textheight]{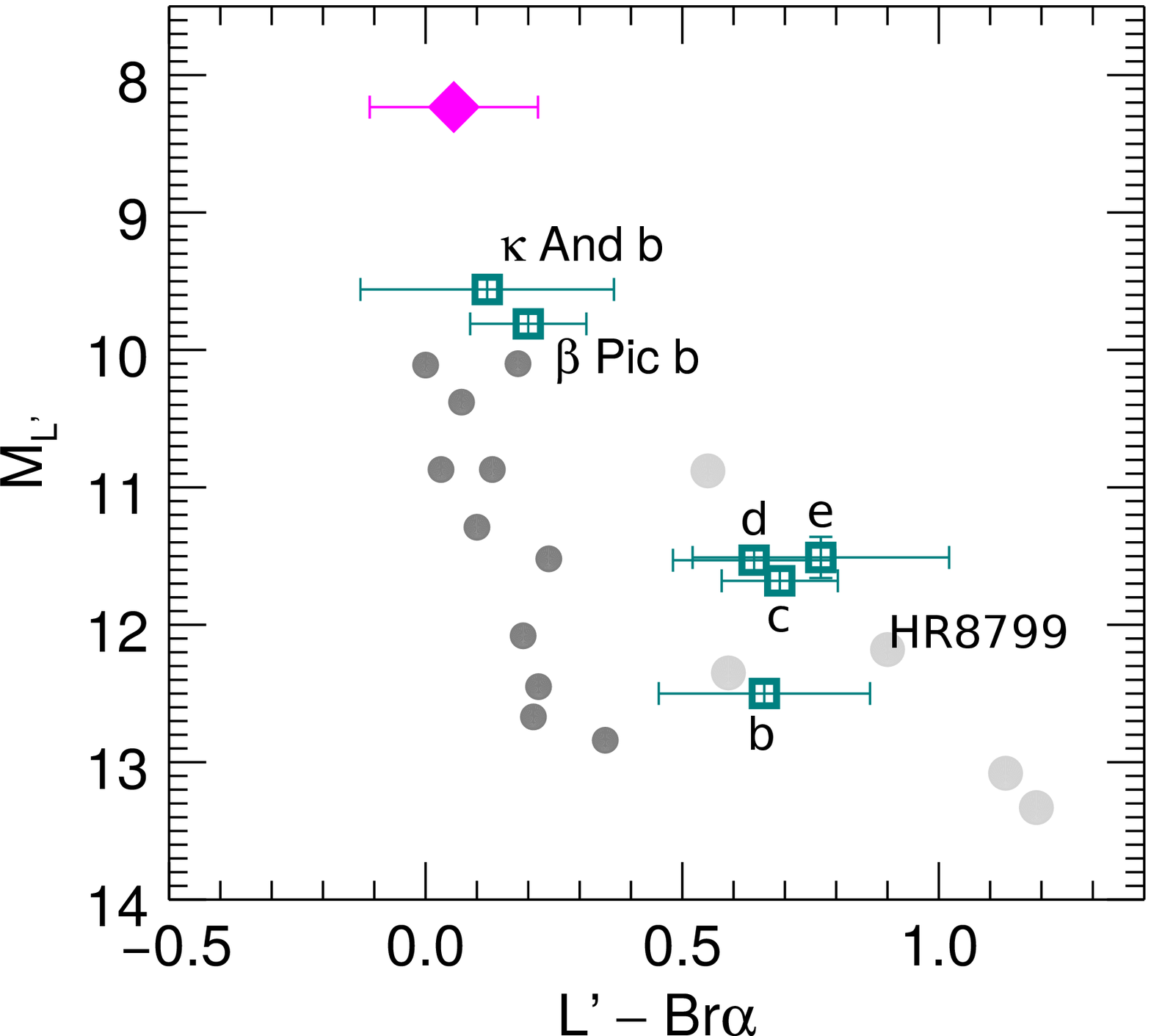}\hfill
\includegraphics[height=0.22\textheight]{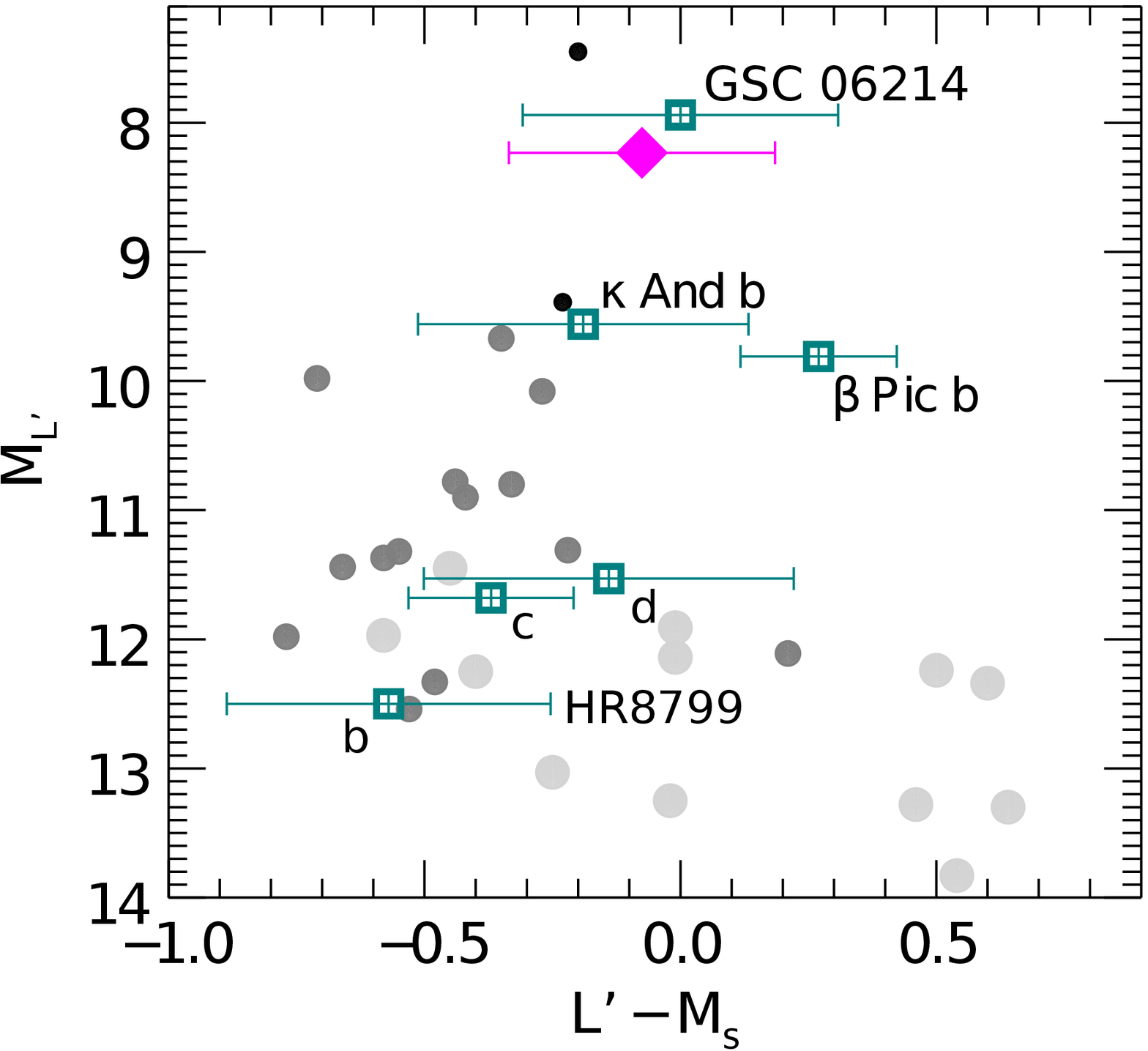}
\caption{Infrared color-magnitude diagrams of \roxsBb\ (symbols as in Fig.~\ref{fig:colorcolor}).}
\label{fig:colmag}
\end{figure*}

Fig.~\ref{fig:colorcolor} shows color-color diagrams featuring the new mid-IR \Lp, \Bralpha, and \Ms-band photometry of \roxsBb.
Like some of the other young low-mass objects (but not, for example, {HR\,8799bcd}), \roxsBb\ takes a rather inconspicuous position with respect to the $J$$-$\Ks\ and \Ks$-$\Lp\ colors of field brown dwarfs, on top of the L-type sequence, close to the M-L spectral type transition. Among young low-mass targets with measured \Ks$-$\Ms\ or \Lp$-$\Bralpha\ color, \roxsBb\ appears as one of the bluest. In these colors, which show no color inversion with temperature (unlike \Lp$-$\Ms, see the right panel of Fig.~\ref{fig:colmag}), this is likely due to its comparably high temperature at a young age compared to the other targets with temperatures in the range of $\sim$1000--2200\,K.

Its position in infrared color-magnitude diagrams (Fig.~\ref{fig:colmag}) reveals a higher luminosity compared to field targets of the same color. 
GSC\,06214\,B is the only other young target in the diagram with a brightness comparable to that of  \roxsBb. However, it has a slightly redder \Ks$-$\Ms\ color, which may be due to circumstellar dust \citep{bai13}. In general, the red appearance of young low-mass objects appears to be a frequent phenomenon \citep{fah16} and has led to suggestions of cloudy atmospheres with vertical mixing and low gravity rather than circumstellar disks, with the caveat that most targets are much older than \roxsB\ ($\sim$2\,Myr) and GSC\,06214\,B ($\sim$10\,Myr), i.e., ages where field targets have a lower probability to show significant signs of circumplanetary material.

\begin{figure}
\centering
\includegraphics[width=0.9\columnwidth]{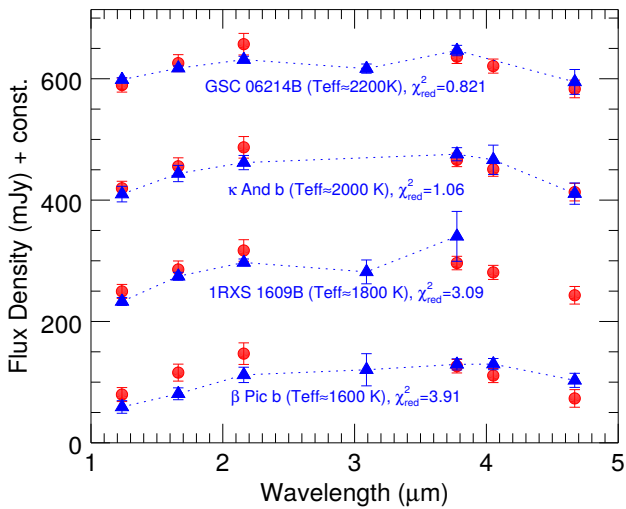}
\caption{Comparison of \roxsBb\ near- to mid-IR photometry (red circles) with other low-mass objects in a similar T$_{\rm eff}$ range (blue triangles and dashed lines). The SEDs have been scaled by a factor that minimizes $\chi^2$ (see Sect.~\ref{sec:results_colors}).}
\label{fig:compare_SEDs}
\end{figure}

A comparison of the spectral energy distribution (SED) of \roxsBb\ with directly imaged objects at a similar temperature ($\beta$\,Pic\,b, $\sim$1600\,K; 1RXS\,1609b, 1800\,K; $\kappa$\,And\,b, 2000\,K; GSC\,06214\,B, 2200\,K) helps to further assess the diversity of SEDs of young low-mass objects and their circumstellar environments. As in \citet{cur13}, we scale the comparison SEDs with a factor whose value is the result of a $\chi^2$ minimization. Best fits and reduced $\chi^2$ values are shown in Fig.~\ref{fig:compare_SEDs}. 
The figure allows us to simultaneously compare all available color information between 1--5\,\mum\ without being biased by the comparably high luminosity of \roxsBb.

We find that the hottest targets, GSC\,06214\,B and $\kappa$\,And\,b, are nominally preferred ($\chi^2_{\rm red}=0.82$ and $\chi^2_{\rm red}=1.1$, respectively) over the cooler objects ($\chi^2_{\rm red}>3$). However, the GSC\,06214\,B\ near-IR slope ($JH$\Ks) appears to be flatter than that of \roxsBb, in contrast to the other three comparison objects, which follow the near-infrared slope closely. This can again be explained by the GSC\,06214\,B\ infrared excess from a circumstellar disk and a slightly hotter temperature than \roxsBb, giving rise to the flatter $JH$\Ks\ slope.

We conclude that the near- to mid-IR SED of \roxsBb\ does not exhibit any distinctive features which would set it formally apart (in a $\chi^2$ test) from other young objects of similar temperature, at least not with the precision of the currently available photometry. However, none of the probed SEDs seems to simultaneously reproduce both the near-IR and mid-IR trends perfectly. A similar conclusion was reached for $\beta$\,Pic\,b \citep{cur13}.

\subsection{Forward modeling constraints on surface gravity, effective temperature, and radius}\label{sec:forwardresults}
Of all the tested models, only BT-COND was not able to return an acceptable fit to the photometric data within the probed parameter range. The lack of fit can be explained by the fact that a smaller parameter range is covered by this model compared to other models wihout atmospheric dust (AMES-COND); low gravities \logg$<$4 were not available in the BT-COND parameter grid for temperatures $<$2600\,K. Because AMES-COND and BT-COND both lead to the rejection of all models with \logg$\ge$4 means that we cannot exclude the possibility that dust-free atmospheres in both frameworks are consistent with low-gravity \logg$\le$3 solutions for the \roxsBb\ atmosphere for temperatures between 1600 and 1800\,K.

Based on 1--5\,\mum\ photometry alone, we find that none of the tested dusty or cloudy atmosphere models should be excluded from being considered as representing the spectral energy distribution of \roxsBb. As is known from previous fitting attempts \citep[e.g.,][]{cur13} and also seen in Table~\ref{tab:fittingresults}, near- to mid-IR photometry alone is only weakly sensitive to surface gravity and our fitting does not strongly constrain this parameter. Effective temperature, however, appears to be better constrained. When assuming that the models describe the atmosphere of  \roxsBb\ sufficiently well, we can exclude the lowest ($<$1700\,K) and highest temperatures ($>$2100\,K) at $>$95\% confidence. 

With the exception of the Burrows models, the best-fit temperatures to the $K$-band spectroscopy are on average higher than the photometry fits, mostly 2500--2900\,K. Higher temperatures from spectroscopy compared to photometrically inferred temperatures were already seen by \citet{cur14b}. The fact that formal significances of the spectroscopic fits are close to 1 for most best fits may suggest an overestimation of the error bars. We thus flag the current results as conservative, but note that future studies with more precise flux calibrated spectroscopy are expected to provide stronger constraints.

Considering the fits to photometry and spectra separately show a preference for models with substantial atmospheric dust/clouds, similar to the results from \citet{cur14b}. The best-fit dust-free models for photometry and those for spectra (considered separately, not jointly) are inconsistent. The AMES/BT-COND model fits to photometry prefer temperatures of $\sim$1700--1800\,K, while fits to the $K$-band spectrum center on 2500--3000\,K. Acceptable fits to the photometry and spectrum for the AMES-DUSTY and BT-SETTL~2015 models are similarly inconsistent. In contrast, a subset of the Burrows, BT-SETTL and BT-DUSTY model fits to photometry also fit the $K$-band spectrum. The union of the Burrows, BT-DUSTY, and BT-Settl model fits cover \teff=2000--2100\,K/\logg=3.4--3.6, and \teff=1700--2000\,K/\logg=4.5--5, \teff=1600\,K/\logg=4.5, \teff=1800\,K/\logg=3.5, respectively.

To assess the influence of mid-infrared photometry on model selection, we compare the best fits that are obtained for the Burrows models when only near-infrared photometry is known to those with 1--5\,\mum\ photometry {(top panel of Fig.~\ref{fig:comparefilters}), as well as spectroscopy fits with and without  1--5\,\mum\ photometry (bottom panel). We find that model} spectra consistent with near-infrared photometry of \teff$\sim$2000\,K objects can vary as much as 50\% in the 2--4\,\mum\ range where the SED is the brightest. This translates into a 25\% uncertainty of the bolometric luminosity. For comparison, the luminosity of spectra consistent with $JH$\Ks\Lp\Bralpha\Ms\ photometry varies by only $\sim$6\%.

When fit simultaneously with the photometric data, the $K$-band spectroscopy appears to dominate the result. While the smallest parameter range is found to fit the combined spectroscopy and photometry, the bottom right panel of Fig.~\ref{fig:comparefilters} demonstrates, using the example of the Burrows models, that they do not fit the spectroscopy and photometry equally well. The same conclusions are reached for the other model families. The on average lower significances of the simultaneous fits reinforce the discrepancy between the photometry and spectroscopy for most model families, as discussed above.

The average inferred radii from the photometry fits range between $\sim$2.4\,\RJup\ and $\sim$3.2\,\RJup. Radii consistent with the spectroscopy fits are bimodal (1.3--2.4\,\RJup\ and 2.6--4.7\,\RJup) where the lower bracket is due to on average higher temperatures of the accepted spectroscopic fits. When compared with stellar evolution models, the photometrically derived radii appear  larger while the radii derived from the spectroscopic data are consistent with the expectations for a 2\,Myr, $M$=9$\pm$3\,\MJup\ object. For example, isochrones from the AMES models suggest radii of 1.9--2.4\,\RJup\ for \roxsBb, very similar radii are derived from the Burrows models \citep{bur01}.

In the AMES framework this combination of radius and mass is equivalent to a surface gravity value of \logg$\approx$3.7, consistent with all fits except the photometry-only fits of the COND models and BT-DUSTY. A low surface gravity is corroborated by empirical results; based on comparison with other young low-mass objects  \cite{cur14a} concluded that \roxsBb\ has a surface gravity of \logg\,=\,3.5--4.5.

\subsection{Atmospheric retrieval modeling results}
\label{sec:molab}

\begin{figure*}
\centering
\setlength{\unitlength}{\textwidth}
\begin{picture}(0.7,0.5)
    \put(0.00,0.00){\includegraphics[width=0.7\textwidth]{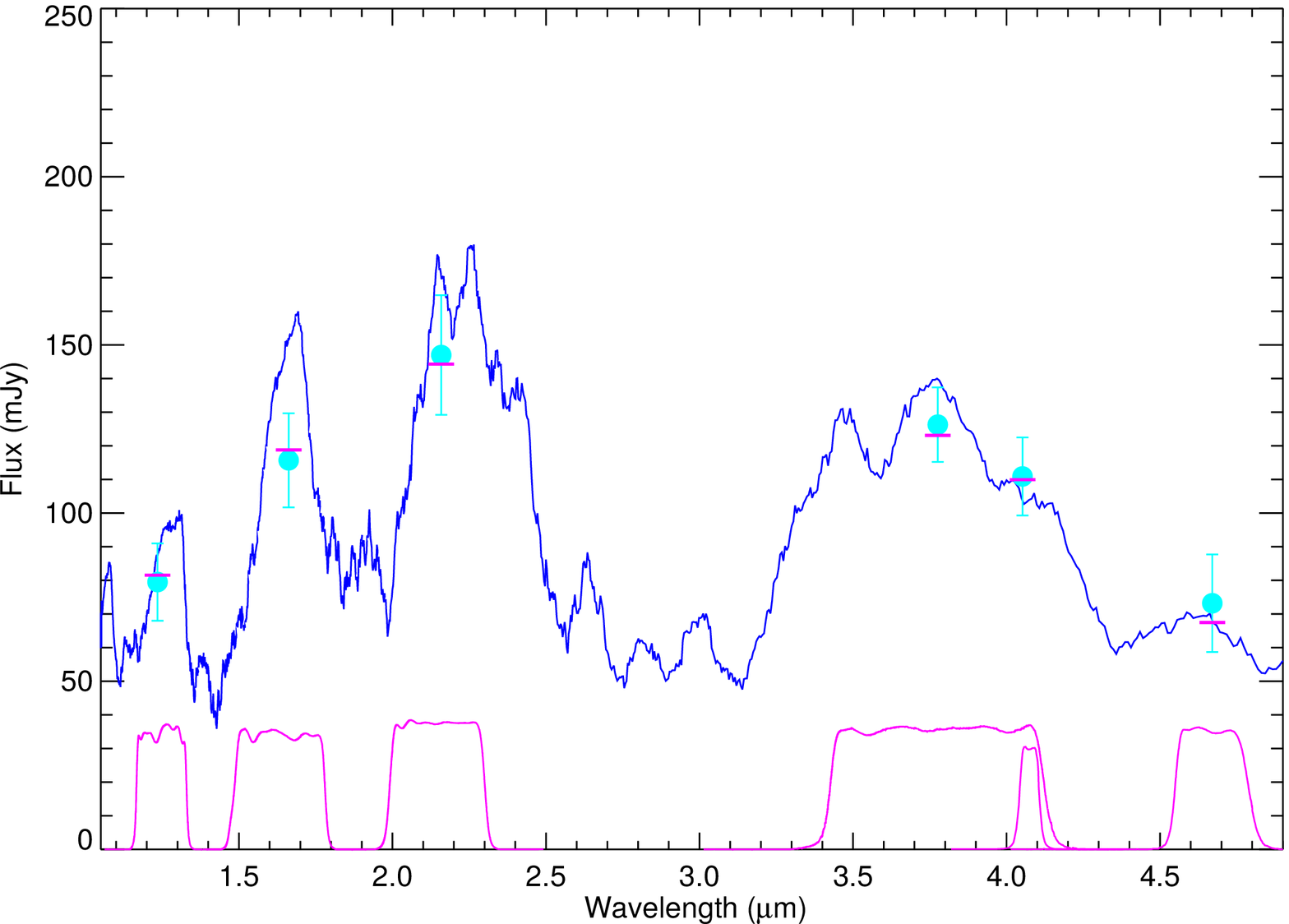}}
    \put(0.39,0.30){\includegraphics[width=0.29\textwidth]{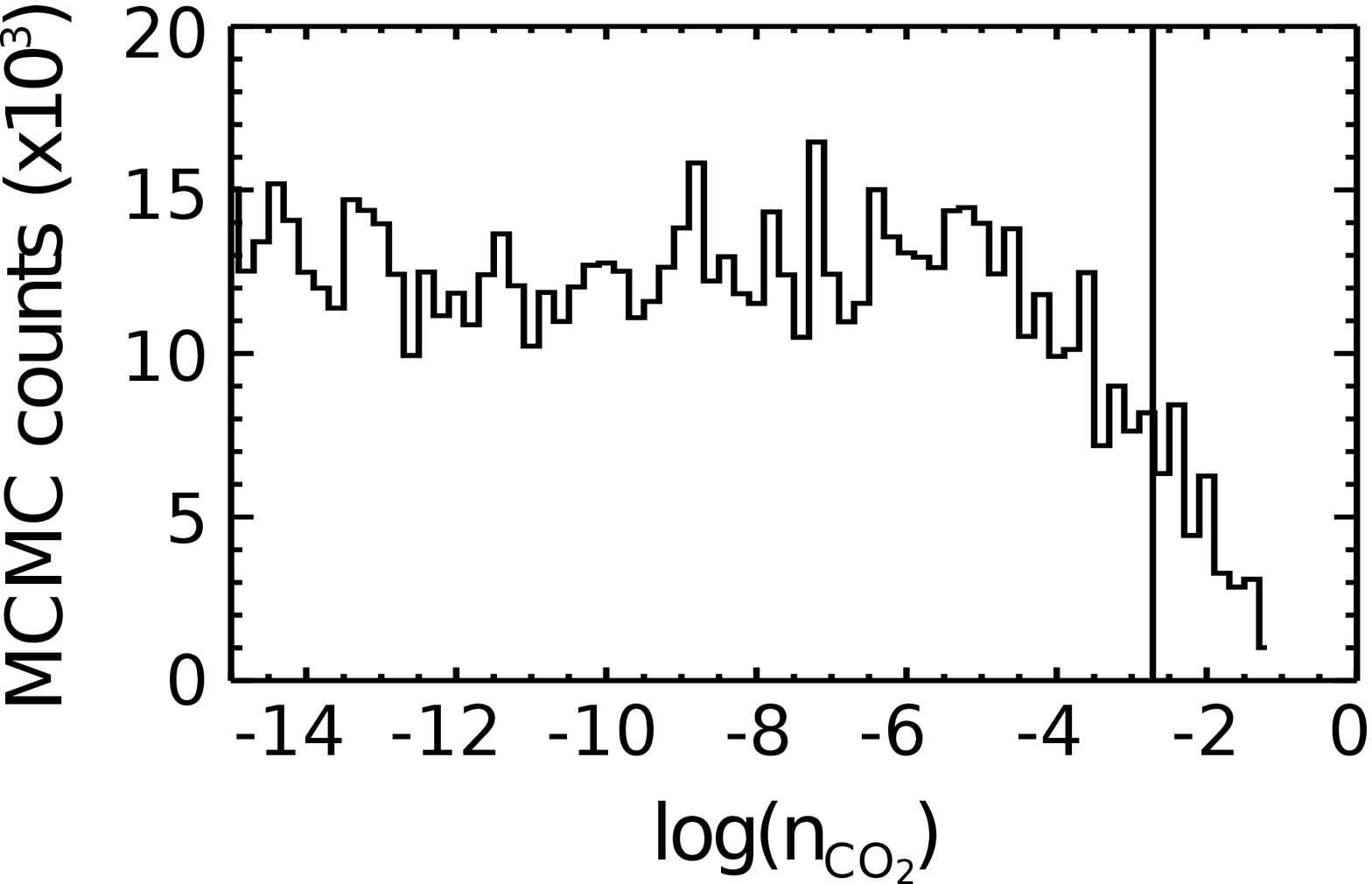}}
\end{picture}
\caption{ Best fit found by the MCMC retrieval run for the photometry of \roxsBb\ ($\chi^2$\,=\,0.35). The observed data is indicated by cyan points with error bars. The blue curve shows a smoothed version of the retrieval emission model and the magenta points represent the model integrated over the transmission function of the photometric bands (magenta curves at bottom) to match the photometry. We note that this is not necessarily the best fit (according to $\chi^2$) since the MCMC parameter states make random jumps, but just the best fit encountered by the Markov chain. The inset shows the histogram of the $\rm CO_2$ molecule number fraction throughout the MCMC run. Values of $\log(n_{\rm CO_2})$ above $-1$ are not permitted by the fitting routine; 95\% of the values are below $-2.7$ establishing the upper limit we adopt. 
}
\label{fig:retrieval}
\end{figure*}

Using ground-based photometry to perform retrieval is challenging for several reasons. First, the photometric bands are \emph{designed} to probe wavelengths with very little absorption from common molecules like water, CO, CO$_2$, and methane. Thus, the retrieval algorithm has very little leverage on the abundances of these molecules in the observed planetary atmosphere. An additional constraint is that broadband photometry in general contains only a small amount of information about the abundances of specific molecules \citep[e.g.,][]{lin12, lin14}. Difficulties in constraining the molecular abundances with broadband photometry also arise from complex degeneracies between molecular absorption, surface gravity, potential clouds, and the pressure-temperature profile of the atmosphere. All of these parameters are strongly correlated or anticorrelated and the degeneracies can very often be broken only by the detection of well-resolved spectral features with a high signal-to-noise ratio \citep[e.g.,][]{lin12}. 
Despite these difficulties, photometry contains some information about the atmospheric composition. Our simplified, cloud-free plane-parallel framework combined with MCMC retrieval (see Sect.~\ref{sec:retrieval}) can reproduce the observed photometry well (Figure~\ref{fig:retrieval}), with the caveat that the fit contains more free parameters than photometric data points. Thus, we are unable to determine robust uncertainties, but in principle we are able to exclude certain classes of models. In this way, we are able to place an upper limit on $\log(n_{\rm CO_2})$ of $-$2.7 (95\% confidence level), within the framework of our retrieval model. The abundances of the remaining major molecular species that we include in our analysis remain unconstrained.
This is not surprising;  even when spectroscopy is available, precise determinations of molecular abundances can be difficult. For example \citet{tod16} use low-resolution 1--1.8\,\mum\ spectroscopy to retrieve abundances of the  $\kappa$\,And\,b\  atmosphere. Using the same fitting parameters as we do (their \emph{case~3}), they derive an upper limit for methane of $\log(n_{\rm CH_4})$\,$<$\,$-2.4$ for $\kappa$\,And\,b (\teff$\approx$2000\,K) and find no constraints for the abundances of $\log(n_{\rm CO})$. Their 95\% upper limit for $\log(n_{\rm CO_2})$ of $-2.2$, however, is consistent with what we find for \roxsBb. Only the water abundance of $\kappa$\,And\,b is fully constrained {($\log(n_{\rm H_2O})$\,$=$\,$-3.5^{+0.4}_{-7.8}$)}.

We note that our upper limit on $\log(n_{\rm CO_2})$ is not strict enough to support or exclude equilibrium chemistry. Equilibrium with other species (such as $\rm H_2O$, CO, $\rm CH_4$) can be represented with a system of equations involving the abundances of multiple chemical species \citep{hen16}. A comparison with equilibrium forward models is difficult without assuming the overall metallicity of the atmosphere and the pressure-temperature structure of the atmosphere (which here are unconstrained). To obtain an estimate, we calculate the equilibrium $\rm CO_2$ abundances for an isothermal atmosphere at 2000\,K with solar abundances of C, O, and He using the open source VULCAN\footnote{\url{https://github.com/exoclime/VULCAN}} code \citep{hen16}. We derive an abundance of $\log(n_{\rm CO_2})< -7.0$ at all pressures, lower than our retrieved upper limit of $\log(n_{\rm CO_2})< -2.7$. The reported upper limit will nevertheless be useful and will serve as a reference value for future spectroscopic atmospheric retrieval results and studies of other targets.

\section{Summary and discussion}\label{sec:discussion}
We present new 3--5\,\mum\ photometry of the directly imaged exoplanet candidate \roxsBb. To analyze the spectra and measure how mid-IR photometry helps to constrain atmospheric parameters, we combine this data with previous near-infrared photometry and spectroscopy, and compare the combined data to atmosphere models and other young low-mass objects.

We find that atmospheric models with either uniformly dusty atmospheres (AMES-DUSTY, BT-DUSTY) or featuring clouds (BT-SETTL, Burrows) are in agreement with the combined 1--5\,\mum\ photometry. As is expected for L-type objects, cloudless models (AMES-COND, BT-COND) generally fit the data  less well, according to their $P$-value. Based on photometry alone, we reject temperatures $>$2100\,K and $<$1600\,K for all models with 95\% confidence (explored parameter range: \teff=1500--3000\,K, \logg=2.5--5.5). While this agrees well with results from the previous fitting of near-IR photometry ($J$--\Lp\ band), which suggested effective temperatures of 1800--2000\,K \citep{cur14b}, the additional photometry helps to significantly narrow the range of acceptable models. Owing to the weak sensitivity of the photometry to surface gravity, no additional constraints can be placed on the  \roxsBb\ gravity beyond the empirical values of \logg$\approx$3.5--4.5 reported by \citet{cur14a}.

Fits of $K$-band spectroscopy of \roxsBb\ \citep{cur14a} result in  average higher best-fit temperatures of mostly 2500--2900\,K, as was seen previously \citep{cur14b}. Models with dust-free (AMES/BT-COND) atmospheres show a particularly strong discrepancy between the photometric and spectroscopic temperature estimates, whereas the dusty and/or cloudy BT-DUSTY, BT-SETTL, and Burrows models each cover a part of the \teff-\logg\ parameter space where both the spectroscopic and photometric fit can be accepted. Nevertheless, we find that the none of the tested models reproduces all the qualitative features equally well in a weighted simultaneous photometric and spectroscopic fit (e.g., the shape of the K-band spectroscopy continuum \emph{and} the 1--5\,\mum\ photometry). Our new analysis thus reinforces the notion by \citet{cur14b}, which was based on 1--3.5\,\mum\ data and a smaller range of atmosphere models, that the  \roxsBb\ infrared photometry and spectroscopy cannot be simultaneously fit by most atmosphere models with satisfactory results. This might point to either insufficiently accurate descriptions of the opacities and temperature structure of planetary atmosphere models or nonstandard element abundances in \roxsBb. The latter branch has been explored for other directly imaged planets by invoking non-equilibrium carbon abundances \citep[e.g.,][]{ske14} or by changing the overall metallicity \citep[][]{ske16,wag16}. While the effective temperature of  \roxsBb\ is likely too high for non-equilibirum chemistry to have a significant effect on the observed infrared properties \citep{mad14}, an increased metallicity may have a noticeable effect on the intrinsic colors and the spectral signature. Metallicity offsets relative to the star may be a sign of the formation of a companion in a circumstellar disk rather than through cloud fragmentation in a stellar formation channel \citep{ske16} and thus open an interesting path to distinguish formation scenarios. Dedicated modeling efforts in future studies will help to constrain this possibility for \roxsBb.

Comparison of the  \roxsBb\ photometric 1--5\,\mum\ SED with previous observations of young  low-mass objects in a similar temperature range return acceptable point-wise agreement with GSC\,06214B and $\kappa$\,And\,b. Nevertheless, none of the targets is simultaneously consistent in near- to mid-infrared colors and luminosity. One of the distinguishing features of \roxsB\ is its very young age of $\sim$2\,Myr (assuming it is a member of the Ophiuchus region). Owing to the currently small number of directly imaged planetary-mass objects, however, it is not clear whether the atmospheric properties of  \roxsBb\   are characteristic of young targets. Furthermore, it is known that young substellar objects, including planetary-mass companions \citep[e.g., GSC\,06241\,B;][]{bai13}, can exhibit circumstellar disks that can have a strong effect on the measured mid-IR SED. It is possible that circumplanetary material might be present around \roxsBb\  given that $\sim$1/3 of all Ophiuchus members harbor circumstellar dust \citep{eri11}. However, additional signs of disks, such as strong emission lines indicative of accretion, are not seen in our data. 

From our cloudless atmospheric retrieval analysis of the  \roxsBb\ 1--5\,\mum\ photometry we infer an upper limit for the abundance of CO$_2$ of $\log n_{\rm CO_2}\!<\!-2.7$. Other probed molecular species (e.g., water, methane, carbon monoxide) or properties (\teff, \logg) are not constrained by the measured filter set. In general, inference of the atmospheric parameters based on six photometric data points make strong constraints on atmospheric properties impossible. An in-depth study will hence require more mid-IR observations as well as infrared spectroscopy over a long wavelength range of reference objects at similar temperatures and ages. In particular, retrieval methods will benefit strongly from new data since parameter constraints are a strong function of the number of data points \citep{lee13}. As such an analysis is beyond the scope of the current paper, we defer the analysis of additional photometry and new spectroscopic measurements of \roxsBb\ to future work (Currie et al.\  in prep.).

\begin{acknowledgements}
We thank the anonymous referee for a critical review of the manuscript and helpful comments. 
We thank Micka\"el Bonnefoy, Sascha Quanz, and Adam Amara for discussing details of the project. We particularly thank Sean Brittain for sharing with us his spectra of HD\,141569. 
This work has been carried out within the frame of the National Centre for Competence in Research PlanetS supported by the Swiss National Science Foundation. S.D.\ acknowledges the financial support of the SNSF. 
This work was supported in part by NSERC grants to R.J. 
The data presented herein were obtained at the W.M.\ Keck Observatory, which is operated as a scientific partnership among the California Institute of Technology, the University of California, and the National Aeronautics and Space Administration. The Observatory was made possible by the generous financial support of the W.M.\ Keck Foundation. The authors wish to recognize and acknowledge the very significant cultural role and reverence that the summit of Mauna Kea has always had within the indigenous Hawaiian community.  We are most fortunate to have the opportunity to conduct observations from this mountain. 
\end{acknowledgements}

\end{document}